\documentclass[aps,epsf,subfigure,twocolumn]{revtex4}
\usepackage{graphicx}
\usepackage{dcolumn}
\usepackage{bm}
\usepackage{amsmath}
\usepackage{amssymb}
\usepackage{subfigure}
\usepackage{color,psfrag,epsfig}

\begin{document}

\newcommand{\bk}{{\bf k}}

\newcommand{\mub}{{\mu_{\rm B}}}
\newcommand{\sD}{{\scriptscriptstyle D}}
\newcommand{\sF}{{\scriptscriptstyle F}}
\newcommand{\sCF}{{\scriptscriptstyle \mathrm{CF}}}
\newcommand{\sH}{{\scriptscriptstyle H}}
\newcommand{\sAL}{{\scriptscriptstyle \mathrm{AL}}}
\newcommand{\sMT}{{\scriptscriptstyle \mathrm{MT}}}
\newcommand{\sT}{{\scriptscriptstyle T}}
\newcommand{\up}{{\mid \uparrow \rangle}}
\newcommand{\down}{{\mid \downarrow \rangle}}
\newcommand{\upt}{{ \langle \uparrow \mid}}
\newcommand{\downt}{{\langle \downarrow \mid}}
\newcommand{\Up}{{\mid \Uparrow \rangle}}
\newcommand{\Down}{{\mid \Downarrow \rangle}}
\newcommand{\bbar}{{\mid \uparrow, 7/2 \rangle}}
\newcommand{\abar}{{\mid \downarrow, 7/2 \rangle}}
\renewcommand{\a}{{\mid \uparrow, -7/2 \rangle}}
\renewcommand{\b}{{\mid \downarrow, -7/2 \rangle}}
\newcommand{\plus}{{\mid + \rangle}}
\newcommand{\minus}{{\mid - \rangle}}
\newcommand{\ex}{{\mid \Gamma_2^l \rangle}}
\newcommand{\braex}{{\langle \Gamma_2^l \mid}}
\newcommand{\LH}{{{\rm LiHoF_4}}}
\newcommand{\LHx}{{{\rm LiHo_xY_{1-x}F_4}}}
\newcommand{\Ht}{{H_{\perp}}}
\newcommand{\Htst}{{H_{\perp}^*}}
\newcommand{\psio}{{\mid \psi_o \rangle}}
\newcommand{\psis}{{\mid \psi \rangle}}
\newcommand{\bpsio}{{\langle \psi_o \mid}}
\newcommand{\barpsi}{{\mid \psi' \rangle}}
\newcommand{\barpsio}{{\mid \bar{\psi_o} \rangle}}
\newcommand{\de}{{{\delta E}}}

\title{The low-$T$ phase diagram of $\LHx$}

\author{M. Schechter$^1$ and P. C. E. Stamp$^{1,2}$}

\affiliation{$^1$Department of Physics and Astronomy, University of
British Columbia, Vancouver, British Columbia, Canada V6T 1Z1 \\
$^2$Pacific Institute for Theoretical Physics, University of British
Columbia, Vancouver B.C., Canada V6T 1Z1.}


\begin{abstract}
The ${\rm LiHo_xY_{1-x}F_4}$ compound is widely considered to be the
archetypal dipolar Quantum Ising system, with longitudinal dipolar
interactions $V_{ij}^{zz}$ between ${\rm Ho}$ spins $\{ i,j \}$
competing with transverse field-induced tunneling, to give a $T=0$
quantum phase transition. By varying the ${\rm Ho}$ concentration x,
the typical strength $V_0$ of $V_{ij}^{zz}$ can be varied over many
orders of magnitude; and so can the transverse field $H_{\perp}$. A
new effective Hamiltonian is derived, starting from the
electronuclear degrees of freedom, and valid at low and intermediate
temperatures. For any such dipolar Quantum Ising system, the
hyperfine interaction will dominate the physics at low temperatures,
even if its strength $A_0 < V_0$: one must therefore go beyond an
electronic transverse field Quantum Ising model. We derive the full
phase diagram of this system, including all nuclear levels, as a
function of transverse field $H_{\perp}$, temperature $T$, and
dipole concentration x. For ${\rm LiHo_xY_{1-x}F_4}$ we predict a
re-entrant critical field as a function of x. We also predict the
phase diagram for x$=0.045$, and the behavior of the system in
magnetic resonance and $\mu$SR experiments.

\end{abstract}

\maketitle

\section{Introduction}

\subsection{The TFQI Model for $\LHx$}
 \label{sec:TFQI}

For at least a decade the $\LHx$ compound has been considered to be
an ideal experimental realization of the well-known 3-dimensional
transverse field Quantum Ising model (TFQIM). According to this
view, at temperatures well below an anisotropy energy $\Omega_0$, it
is described by the Hamiltonian:
\begin{equation}
  H =  - \sum_{i,j} V_{ij}^{zz} \tau_i^z  \tau_j^z - \Delta_0
\sum_i \tau_i^x \, ,
 \label{Q-Ising}
\end{equation}
where $\vec{\tau}_j$ is a Pauli vector describing a two-level
effective electronic spin at spatial position ${\bf r} = {\bf r}_j$,
$V_{ij}^{zz}$ is a longitudinal inter-spin interaction, with
nearest-neighbor strength $U_0$ which, depending on the dilution x,
can have either ferromagnetic or a frustrating character; and the
'transverse field' term $\Delta_0 \sum_i \tau_i^x$ is controllable
externally (usually by applying a transverse magnetic field). The
most distinctive feature of the TFQI Model (\ref{Q-Ising}), which is
central to the whole field, is the competition between $V_0$, which
tries to order the system, and $\Delta_0$, which causes quantum
fluctuations out of the ordered state. At $T=0$ one expects a
quantum phase transition between ordered and quantum disordered
states when $\Delta_0/V_0 \sim 1$, and this is probably the simplest
theoretical example of a quantum phase transition. The apparent
confirmation of this 'quantum critical' picture for $\LHx$ has lent
considerable importance to the experiments on this system.

The main arguments in favor of this picture for $\LHx$ are as
follows:

(i) The strong crystal field Ho single-ion anisotropy yields an
Ising doublet ground state, with a crystal field Hamiltonian
yielding an appreciable $\Delta_0$ at small $H_{\perp}$. The
dominant inter-Ho spin-spin interaction is dipolar, with strength
$V_0({\rm x})=\sum_j \langle V_{ij}^{zz} \rangle \sim \alpha {\rm
x}$, with $\alpha \sim 1$ in Kelvin units. Thus when x$=1$ one
expects a dipolar-ordered ferromagnetic phase below $\sim 1$K, which
is observed; it exhibits both classical and quantum phase
transitions to the paramagnetic phase\cite{BRA96}.

(ii) In $\LHx$ the magnetic Ho ions and the non-magnetic Y ions have
very similar atomic volumes; dilution of the Ho by Y is then
possible with negligible distortion of the lattice.  This dilution
weakens the interactions and introduces randomness and frustration;
very different physical regimes can then be
studied\cite{REY+90,Ros96}; see Fig. 1 in Ref. \cite{REY+90}. In
particular, one expects a low-$T$ spin-glass phase at small x, below
a transition temperature $T_c \sim \alpha $x. At x$=0.167$ a
spin-glass phase is found\cite{WER+91,WBRA93} at low $T$ and
$H_{\perp}=0$, with a crossover to the paramagnetic phase at higher
$T$ and $H_{\perp}$. At x$=0.44$ the tunneling of domain walls in
the ferromagnetic phase was found\cite{BRA01} and differences
between quantum and classical annealing protocols were
observed\cite{BBRA99}.  At x$=0.045$ the system shows a peculiar
narrowing of the spin fluctuation spectral width as temperature is
decreased\cite{GPRA02}, described as ``anti spin-glass'' behavior.

(iii) For extreme dilution one expects single Ho ion behavior. In
experiments at x$=0.002$, hysteresis loops of the magnetization due
to single spin tunneling are observed\cite{GWT+01} (co-tunneling of
pairs of spins was also observed at x$=0.002$, showing that
interaction effects cannot be neglected even at this dilution
\cite{GTB03,BGW+04}).

\vspace{2mm}

Thus, according to these arguments, a TFQI model like
(\ref{Q-Ising}) should describe $\LHx$ for all x, provided $kT,
\mu_BH_{\perp} \ll \Omega_0$; and as such, $\LHx$ should be a model
system for all dipolar magnets. However we argue in this paper that
the $\LHx$ system (and by implication, many other dipolar magnets)
need to be described in a quite different way. There are two main
problems with the simple TFQI picture, both noted and analysed in
ref.\cite{SS05}. These are

(a) {\it Hyperfine Interactions}: The on-site Ho hyperfine
interaction $A_0$ is not small - in fact even at x$=1$, $A_0 \sim
V_0$, and for x$ \ll 1$, the hyperfine interaction is overwhelmingly
dominant! A few experimental papers have heeded this point,
remarking (i) that even the x$=1$ phase diagram, near the $T=0$
ferromagnetic-paramagnetic transition, is modified by the hyperfine
interaction\cite{BRA96}; and (ii) that the nuclear spin bath,
considered now as a quantum environment\cite{PS00}, should strongly
affect the Ho spin dynamics near this quantum critical
point\cite{RPJ+05,RJP+07}. However we shall show here that the
effect of nuclear spins on dipolar magnets is much more profound
than this, {\it even when the hyperfine interaction is quite weak}.
This very surprising result means that one must reconsider the
application of the TFQI Hamiltonian to a large variety of systems,
hitherto analysed without reference to the hyperfine couplings.

(b) {\it Transverse Dipolar Interactions}: When x$ \neq 1$ these
interactions add a quite large contribution to the transverse field
- to quantitatively understand the phase diagram one then needs to
include them\cite{SS05,SL06} (see also ref.\cite{GRAC03}), both in
the spin-glass and in the ferromagnetic
regimes\cite{SS05,SL06,TGK+06,SSL07,Sch06}.

\subsection{An ENQI Model for Dipolar Ising Magnets}
 \label{sec:ENQI-intro}

To properly treat the physics of Quantum Ising systems, we have to
recognize that the use of a simple parameter $\Delta_0(H_{\perp})$,
introduced a long time ago by experimentalists as a convenient way
of defining an effective transverse field acting on the Ising spins,
is actually misleading. Because of the nuclear spins, the true
effective transverse field in a quantum Ising system is very
different from $\Delta_0$; moreover it depends on the actual nuclear
spin state of the system.

In what follows we will derive a theoretical framework with the
nuclear spins included from the beginning. The system is described
at low energies in terms of 'electronuclear' complexes which
interact via renormalized dipolar interactions. In its general form
[see Eq.(\ref{mIsingH-eff}) just below] this 'Electronuclear Quantum
Ising" (ENQI) Hamiltonian includes all the nuclear spin levels.
However at very low $T$ or for small x, we can use a much simpler
Hamiltonian referring only to the lowest electronuclear doublet, and
this takes the form
\begin{equation}
H = - \sum_{i,j} \tilde{V}_{ij}^{zz}(H_{\perp}) s_i^z  s_j^z -
\tilde{\Delta}(H_{\perp}) \sum_i  s_i^x \, ,
 \label{IsingH-eff}
\end{equation}
where now $\hat{s}_j$ operates only on the single electronuclear
doublet involving the nuclear states with $I_z = \pm I$. Now this
simplified model looks like the standard TFQI model in
(\ref{Q-Ising}), but it behaves very differently - both
$\tilde{V}_{ij}^{zz}(H_{\perp})$ and $\tilde{\Delta}(H_{\perp})$ are
renormalized from their original values in (\ref{Q-Ising}), and they
depend strongly on $H_{\perp}$ [in the case of
$\tilde{\Delta}(H_{\perp})$, this dependence is radically different
from that in the original parameter $\Delta_0 (H_{\perp})$]. The
strength and behavior with field of these variations depends
crucially on the strength $A_0$ of the hyperfine interaction;
moreover, as noted above, we must use this ENQI model at low $T$
even when the hyperfine coupling $A_0 \ll V_0$, which is more
typical for a general anisotropic magnet.

More generally, when $kT$ is not small compared to the splitting
between nuclear levels, we must define a set of $2I + 1$
electronuclear 'pseudospins' (each of which are spin-$1/2$ doublets)
labelled by quantum numbers $m = I, I-1, ... -I$, an occupation
number $n_{im}$ for the occupation of a given pseudospin on site
$i$, and a set of pseudospin operators $\hat{s}_{im}$ and pseudospin
energies $\epsilon_m$; we have the general ENQI Hamiltonian
\begin{eqnarray}
H = &- \sum_{i,j,m,m'} \tilde{V}_{ij,m,m'}^{zz}(H_{\perp}) n_{im}
n_{jm'} s_{im}^z
s_{jm'}^z \nonumber \\
&- \sum_{i,m} n_{im} [\epsilon_m + \tilde{\Delta}_m(H_{\perp})
s_{im}^x] \, ,
 \label{mIsingH-eff}
\end{eqnarray}
where the $\tilde{V}_{ij,mm'}^{zz}(H_{\perp})$ represent
interactions between pseudospins $m,m'$ on different sites $i,j$,
and the transition matrices $\tilde{\Delta}_m$ only operate on
individual pseudospins, ie., within the space of each electronuclear
doublet on a given site. We can think of a set of $2I+1$ independent
quantum Ising systems, each having a different 'transverse field'
$\tilde{\Delta}_m$, which however can interact via the longitudinal
fields $\tilde{V}_{ij,mm'}^{zz}(H_{\perp})$.

In disordered dipolar-coupled spin systems, one must also add a term
which describes the random transverse couplings in the system. Its
detailed form is given in section \ref{Sec:transverse}, and its
quantitative effects are discussed in section \ref{Sec:offD}.

In this paper we concentrate on the $\LHx$ system, for which precise
results and experimental predictions can be established for the
phase diagram, so it can be used as a test case. The effective
Hamiltonian is strictly applicable to systems where $A_0, V_0 \ll
\Omega_0$, in the regime where $T, \mub \Ht \ll \Omega_0$. This
approach enables (i) illumination of the relevant physics of the
$\LHx$ system (ii) generalization to other systems, e.g. systems in
which $A_0 \ll V_0$ (see Sec.~\ref{Sec:offD}) (iii) constructing a
framework for the treatment of dynamical properties. However, in the
$\LHx$ system the condition $A_0, V_0 \ll \Omega_0$ is not that well
satisfied, and the condition $T, \mub \Ht \ll \Omega_0$, while is
satisfied in the whole relevant phase diagram at low x, is not
satisfied near criticality at large concentrations. For this reason,
and since single ion properties dictate much of the physics in the
$\LHx$ system, we also use exact diagonalization of the Ho
electronuclear spin states. This enables us to give quantitative
predictions regarding the single ion characteristics, and with the
use of mean field approximation, to predict the form of the phase
diagram for all x.

Most of the results here are new. We analyze in detail the form of
the electro-nuclear states of the single Ho ion as function of
$\Ht$, and its consequences in terms of entanglement entropy and
magnetic resonance experiments. We show that the peculiar crystal
field Hamiltonian of $\LHx$ results in a well defined Ising system
even at high transverse fields (where Ising symmetry is usually
destroyed). We obtain a general effective Hamiltonian valid for
thermodynamic properties, incorporating all $16$ low energy states,
therefore generalizing the treatment in Ref.\cite{SS05} to the
regime $A_0 < T \ll \Omega_0$. We give a discussion of the phase
diagram for general concentration x, temperature $T$, transverse
field $H_{\perp}$ and hyperfine coupling $A_0$. With relevance to
general magnetic systems we show that the hyperfine interactions
dominate the physics at low T even when $A_0 \ll V_0$. By comparing
the phase diagrams at x$=0.045$ and x$=0.167$ we predict a novel
reentrance of the crossover transverse field between the quasi-SG
and PM phases at low $T$ as function of x, resulting from the
interplay between the hyperfine and off-diagonal dipolar
interactions. We then add a novel perspective to the unsolved
question of the nature of the low-T phase at x$=0.045$. Finally, we
discuss further experimental consequences of our results.

The paper is organized as follows: In Sec.~\ref{Sec:Ho} the various
terms in the microscopic Hamiltonian for $\LHx$ are introduced and
quantified, and single ion properties are analyzed. In
Sec.~\ref{Sec:transverse} the full low energy effective Hamiltonian
is derived, including the transverse hyperfine interactions and the
off-diagonal terms of the dipolar interaction. In
Sec.~\ref{Sec:offD} we obtain the phase diagram of the $\LHx$ system
at different dilutions; we obtain quantitative agreement with the
experimental phase diagram at x$=0.167$, make predictions regarding
the phase diagram at x$=0.045$, and discuss the nature of the low
temperature phase. In Sec.~\ref{experiments} we suggest experiments
that can directly check our theory, and in Sec.~\ref{sec:conc} we
state our conclusions. Some details regarding the derivation of the
effective Hamiltonian and the calculation of the phase diagram in
mean field are deferred to appendices.

\section{Interactions in the $\LHx$ system}
 \label{Sec:Ho}

In this section we give the quantitative form of the $\LHx$
Hamiltonian, which is a sum of crystal field\cite{GWT+01,CHK+04},
Zeeman, inter-Ho, and hyperfine interaction terms:
\begin{equation}
H = H_{\rm cf} + H_{\rm Z} + H_{\rm int} + H_{\rm hyp} \, .
 \label{generalH}
\end{equation}
Note that we have dropped: (i) the spin-phonon interaction,
important for spin relaxation\cite{GWT+01,GTB03,BBG+06}; (ii)
hyperfine interactions between the Ho ion and other nuclear species
(F, Li) as well as with Ho nuclei on nearby sites; and (iii) the
nuclear Zeeman couplings. None of these terms have an appreciable
effect on the phase diagram of $\LHx$. Note however that they will
be crucial for the low-$T$ Ho spin dynamics, since even very small
hyperfine terms can strongly affect relaxation dynamics and
decoherence in the low-T quantum regime\cite{PS96,PS98}, where
phonon relaxation is also important in strong transverse
fields\cite{PS96,TS04}.

\subsection{TFQIM terms}
 \label{sec:TFQIM}

Let us first consider the terms which feed directly into the TFQIM
Hamiltonian (\ref{Q-Ising}), ie., the terms $H_{\rm cf}, H_{\rm Z}$,
and $H_{\rm int}$. For $\LHx$ these are given in turn by:

(i) The 'crystal field' term $H_{\rm cf}$ includes the single-ion
crystal field and spin-orbit terms\cite{GWT+01,CHK+04}. Because of
the very strong spin-orbit coupling, $J$ is a good quantum number
for the Ho ion with $J=8$. A crystal field term of form ($J_+^4 +
J_-^4$) strongly mixes states with $J_z$ differing by $\pm 4$
\cite{HJN75,GWT+01}, and a strong $J_z^2$ term severely distorts the
level spacing. There are other terms as well - for computations in
this paper we will use a form written in terms of the usual Stevens
operators as\cite{CHK+04,RPJ+05}
\begin{equation}
H_{\rm cf} = \sum_{l=2,4,6} B_l^0O_l^0 + B_6^4 O_6^4(S) +
\sum_{l=4,6} B_l^4 O_l^4(C)
 \label{LiHo-cf}
\end{equation}
with values assumed to be \cite{CHK+04}
\begin{eqnarray}
B_2^0 = -0.696, \;\;\;\;\;\;\;\;\;\;\; B_4^0 = 4.06 \times 10^{-3}, \nonumber \\
B_6^0 = 4.64 \times 10^{-6}, \;\;\;\;\; B_4^4(C) = 4.18 \times
10^{-2},
\nonumber \\
B_6^4(C) = 8.12 \times 10^{-4}, \;\;\;\; B_6^4(S) = 1.137 \times
10^{-4},
 \label{Bvalues}
\end{eqnarray}
from which we see that the $O_6^4(C)$ term also has a non-trivial
effect, bringing in a $J_z^2 (J_+^4 + J_-^4)$ term (note that to
properly judge the relative importance of coefficients $B_l^m$ and
$B_{l'}^m$ with $l \neq l'$, we should directly compare $J^l B_l^m$
and $J^{l'}B_{l'}^m$, and {\it not} $B_l^m$ and $B_{l'}^m$).

The ground state is an Ising doublet, with states denoted here by
$\up$ and $\down$, which mix states with $J_z=\pm7,\pm3,\mp1,\mp5$.
The first excited state $\ex$ is roughly $\Omega_0=10.5K$ above the
ground state doublet, and is a mixture of $J_z=6,2,-2,-6$. The other
$14$ states are much higher in energy, and the total span of the
$J=8$ manifold is roughly $\Omega_f=500K$\cite{HJN75}.

\vspace{2mm}

\begin{figure}
\includegraphics[width = \columnwidth]{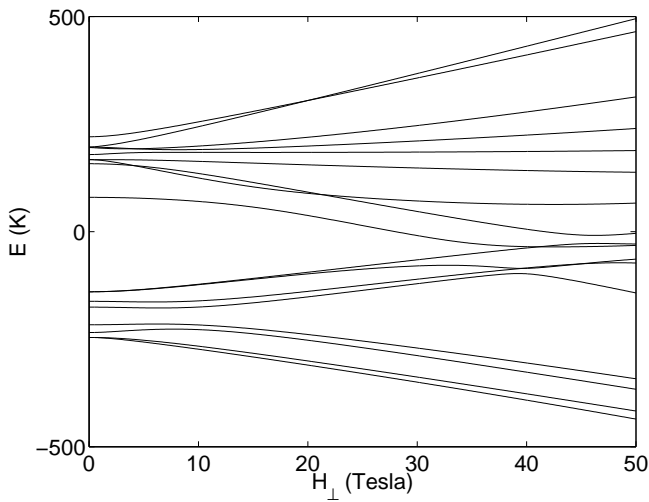}
 \caption{Energy of the $17$ electronic states of a single Ho ion
in $\LHx$, ignoring all inter-Ho interactions, as a function of
transverse field $\Ht$. The Ising-like character is well maintained
until $\Ht \approx 20$T. }
    \label{alllevels}
\end{figure}

(ii) The Zeeman coupling to the Ho spins is given by the usual form
\begin{equation}
H_{\rm Z} = - \sum_i g_J \mub \vec{H} \cdot\vec{J_i}
 \label{H-Z}
\end{equation}
with $g_J=5/4$. We are particularly interested in the effect of a
transverse field $H_{\perp} \ll \Omega_0/\mub$, which induces a
coupling $\Delta_0$ between the two Ising ground states in second
order perturbation theory via the state $\ex$. Thus, for small
fields $\Delta_0 \propto H_{\perp}^2$; by putting in the numbers one
finds
\begin{equation}
\Delta_0 (H_{\perp}) \sim 9 (\mu_BH_{\perp})^2/\Omega_0 \approx 0.4
{\rm T}^2[{\rm K}]
 \label{delta-0}
\end{equation}
in Kelvin units (see e.g. Figs. 1,2 in Ref.~\cite{CHK+04}). At
larger fields, $H_{\perp} \approx \Omega_0/\mub \approx 2$T,
perturbation theory breaks down, $\ex$ mixes strongly with $\up$ and
$\down$\cite{CHK+04}, and $\Delta_0$ is approximately linear in
$H_{\perp}$. An important feature of $\LHx$ is that $\Omega_f \gg
\Omega_0$. Thus, the system stays Ising-like even when $H_{\perp} >
\Omega_0/\mub$, deep inside the paramagnetic regime (see
Fig.\ref{alllevels}). This contrasts with most other anisotropic
dipolar systems, which are dominated by easy-axes terms, so that the
same energy scale dictates the anisotropy and the quantum
fluctuations, and in the quantum phase transition regime $\Ht
\approx H_{\perp}^c$ there is no real Ising character.

\vspace{2mm}

(iii) The inter-Ho ion spin-spin interactions depend strongly on the
Ho concentration x. They have the general form
\begin{equation}
 H_{\rm int} = - \sum_{ij} U_{ij}^{\alpha \beta} \,
J_i^\alpha J_j^\beta .
 \label{U-int}
\end{equation}
Experiments\cite{MJH84} and theoretical analysis \cite{CHK+04} both
show that $U_{ij}^{\alpha \beta}$ is dominated by the dipolar
interaction, ie.,
\begin{subequations}
\begin{eqnarray}
U^{\alpha \beta}_{ij} = \frac{U_0 {\cal R}_{ij}^{\alpha
\beta}}{J^2},\\
{\cal R}_{ij}^{\alpha \beta} = {\cal V}_c \frac{|{\bf r}_{ij}|^2
\delta_{\alpha \beta} - 3 r_{ij}^{\alpha} r_{ij}^{\beta}}{|{\bf
r}^{ij}|^5} .
 \label{UR}
\end{eqnarray}
\end{subequations}
Here the strength of the nearest-neighbor dipole-dipole interactions
between the spins ${\bf J}_i, {\bf J}_j$ is
\begin{equation}
U_0 = \frac{\mu_o}{4 \pi} \frac{g^2_J \mu^2_B J^2}{{\cal V}_c}
 \label{V-0}
\end{equation}
where ${\cal V}_c$ is the unit cell volume; the unit cell size is
(1,1,2.077) in units of $\tilde{a}=5.175 \AA$, with, when x$=1$,
four Ho ions per unit cell, at positions (0,0,0), (0, a/2, c/4),
(a/2, a/2, -c/2) and (a/2, 0, -c/4); and ${\bf r}_{ij} = {\bf r}_i -
{\bf r}_j$.  When x$ < 1$ some of the Ho ions are substituted by Y
ions, and the couplings acquire a random distribution whose
character depends strongly on x. The typical value $\bar{U}_0$ of
the nearest-neighbor coupling then becomes roughly $\bar{U}_0 \sim
{\rm x}U_0$. For $\LHx$, $U_0 \sim 0.3$K. Note however that the
energy $V_0$ characterizing the total effect of the longitudinal
dipolar interactions is somewhat larger than $\bar{U}_0$, since as
noted in the introduction, $V_0({\rm x})=\sum_j \langle V_{ij}^{zz}
\rangle$. Thus the strength of $V_0$ depends on how the spins are
arranged. It can be estimated from the ordering temperature, and
typically $V_0/\bar{U}_0 \sim 3-5$. One can see departures from
linearity in x; for example in the diluted system, even for rather
small x, close pairs and even triplets can dominate certain
properties. There are also antiferromagnetic exchange interactions
between the Ho ions, which for x$=1$ were measured to be about half
of the nearest neighbor dipolar interaction\cite{MJH84}. Therefore,
the exchange interactions have little quantitative significance even
for the undiluted $\LH$\cite{CHK+04}, and are completely negligible
for x$ \ll 1$.

\vspace{3mm}

If we now take these 3 terms and truncate the Ho ions to their
lowest doublet, we get back the TFQIM in (\ref{Q-Ising}), which
predicts a quantum phase transition for x$=1$ at a transverse field
where $\Delta_0 \sim V_0$, ie., at $H_{\perp} \approx
3$T\cite{BRA96,CHK+04}. In fact the actual transition happens at
$H_{\perp}^c({\rm x}=1) = 4.9$T\cite{BRA96,CHK+04}, which is the
first sign that there is something wrong with this naive picture. To
see what is going on we now have to include the hyperfine coupling.

\subsection{Hyperfine interactions}
 \label{sec:hyperfine}

The hyperfine coupling of a single Ho atom with its own $I = 7/2$
nuclear spin gives a term
\begin{equation}
H_{\rm hyp} = A_J \sum_i \vec{I_i} \cdot \vec{J_i} \, ,
 \label{H-hyp}
\end{equation}
with $A_J=0.039$K\cite{GWT+01}. Here we ignore quadrupolar terms as
well as the hyperfine interactions to all other species (Li,F, and
other Ho ions); both are an order of magnitude smaller\cite{SVM+05},
and hardly influence the phase diagram.

At low energies, in the lowest doublet states $\up,\down$, the
longitudinal hyperfine term $H_{\rm hyp}^{\|} = A_J I^z J^z$ splits
each electronic state into an eightfold multiplet of nearly
equidistant levels, with separation $\sim 205 mK$\cite{GWT+01}
between adjacent levels, ie., we can write
\begin{equation}
H_{\rm hyp}^{zz} \sim \omega_0 \tau_z I_z
 \label{H-long}
\end{equation}
where $\hat{\tau}$ operates on the electronic doublet and for
$\LHx$, $\omega_0 \sim 205~mK$; this corresponds to a spin moment
$\langle J_z \rangle \sim 5 \mu_B$ for the lowest doublet.

\begin{figure}
\includegraphics[width = \columnwidth]{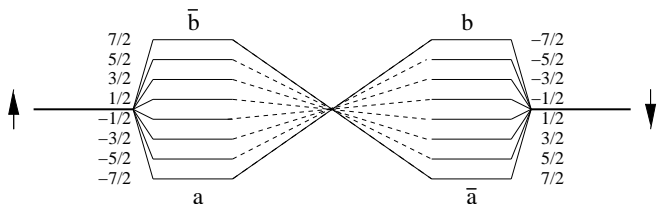}
 \caption{Splitting of the electronic low energy doublet ($\uparrow$ and
$\downarrow$) by the longitudinal hyperfine interaction. The ground
state doublet states $a$ and $\bar{a}$ have a definite and opposite
nuclear spin, $\pm 7/2$. A transverse magnetic field $\Ht$ couples
states with the same nuclear spin, as shown by the dashed lines. }
    \label{fignuclearsplitting}
\end{figure}

One can see without any reference to experiments on the phase
diagram that the TFQI model cannot possibly be right at low
transverse fields, using Fig.\ref{fignuclearsplitting}.  The lowest
energy Ising doublet states $a$, $\bar{a}$ have a definite nuclear
spin ($I_z=-7/2$ for $\up$, and $I_z=7/2$ for $\down$) when $\Ht =
0$. A transverse magnetic field couples $a \equiv \a$ to $b \equiv
\b$ and $\bar{a} \equiv \abar$ to $\bar{b} \equiv \bbar$, and cannot
induce quantum fluctuations between the relevant Ising doublet
ground states at all, but only renormalize their effective spin.
Only the {\it transverse} hyperfine term $H_{\rm hyp}^{\perp}=A_J
(I^+ J^- + I^- J^+)/2$ can change $I_z$, and allow transitions
between the Ising doublet states; but this hardly operates if $\mu_B
H_{\perp} \ll \Omega_0$.

Thus hyperfine interactions must be included in any truncation of
the system to a low-energy Hamiltonian. Their general effect is to
suppress quantum effects at low fields. We shall see that they are
important even when $A_0 \ll V_0$ (note that the simple argument
above, showing the importance of the hyperfine effects, makes no
reference to the strength of these interactions!).

\subsection{Single Ho ion - exact results for low energies}
 \label{sec:singleHo}

\begin{figure}
\includegraphics[width = 4.2in]{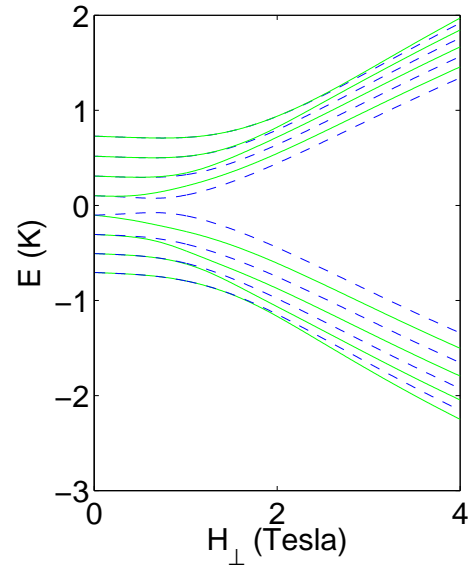}
 \caption{The $16$ lowest electro-nuclear energy levels of $\LHx$,
plotted as a function of transverse magnetic field $\Ht$. The zero
of energy is defined by the (field dependent) mean of levels $8$ and
$9$. Symmetric/antisymmetric states are plotted in solid
green/dashed blue lines. }
    \label{fig16levels}
\end{figure}

For $\Ht \gg \Omega_0/(\mub \langle J_z \rangle)$, $H_{\rm
hyp}^{\perp}$ mixes appreciably electronuclear states with different
values of $I_z$. This is best seen by performing an exact
diagonalization of the full single Ho Hamiltonian $H=H_{\rm
cf}+H_{\rm Z}+H_{\rm hyp}$ in the $136$ eigenfunction space ($17$
crystal field $\times 8$ nuclear states). In Fig.~\ref{fig16levels}
we plot the spectrum of the lowest 16 levels, corresponding to the
electronic ground state doublet, as function of $\Ht$. Most
generally, each of the $16$ states can be written in the form
$\sum_{M,m} \alpha_{M m} \mid M, m \rangle$, where $M(m)$ denote the
z component of the electronic (nuclear) spin. Plotted in solid line
are symmetric eigenstates, with $\alpha_{M m} = \alpha_{-M -m}$, and
in dashed line anti-symmetric eigenstates, with $\alpha_{M m} = -
\alpha_{-M -m}$.

At low fields, the electro-nuclear entanglement is strong, and
states are given, to a good approximation, by the form in
Eq.(\ref{pmstates}). One can then define the splitting between each
pair of time-reversed states by $\tilde{\Delta}_{m}$, which are
plotted in Fig.\ref{figsplitting}. We find\cite{SS05}, as we expect,
that $\tilde{\Delta}_{7/2}$ is small up to $\Ht \approx 2T$, at
which point $\upt \mub H_{\perp} \ex \approx \Omega_0$.
$\tilde{\Delta}_{m}$ increases more rapidly as $|m|$ decreases,
simply because for smaller $|m|$, transitions between the 2
low-energy time-reversed states can be achieved by lower orders in
perturbation theory in $H_{\rm hyp}^{\perp}$; one then sees
appreciable coupling at lower $H_{\perp}$. As $\Ht$ continues to
increase, $\tilde{\Delta}_m$ increases rapidly and eventually
saturates at a field $\Htst(m)$. Note however that the spectrum in
Fig.\ref{fig16levels} is not symmetric. This is because tunneling
between the lower pairs is allowed via the state $\ex$ at energy
$\Omega_0$, whereas tunneling between the upper pairs must involve
the higher excited states, at energy $E_{CF}$ higher than the lowest
states. Consider, eg., the pairs $\mid \uparrow, -1/2 \rangle, \mid
\downarrow, 1/2 \rangle$ and $\mid \uparrow, 1/2 \rangle, \mid
\downarrow, -1/2 \rangle$. The first pair has a finite matrix
element in second order perturbation $\langle \uparrow, -1/2 \mid
H_x J_x \ex \braex I_x J_x \mid \downarrow, 1/2 \rangle$, which is
first order in $\Ht$, and gives a low-field splitting $\propto H_x
I_x/\Omega_0$; the second pair has a term of similar form which
however passes via the states in the multiplet $J_z={8,4,0,-4,-8}$,
and so gives a low-field splitting $\propto H_x I_x/E_{CF}$, roughly
an order of magnitude smaller.

For $\Ht \gtrsim \Htst(m)$ different values of $m$ are well mixed,
the electron and nuclear spins get disentangled, and the spectrum
separates to two groups of 8. For $\Ht \gtrsim \Htst(7/2)$ the
eigenstates can be approximated by $\mid \psi_I \rangle \mid \psi_J
\rangle$. The electronic state hybridizes strongly the level $\ex$
with the ground state doublet. The states in the bottom group are
approximately symmetric with respect to the electronic degrees of
freedom, i.e. have $\alpha_{M m} \approx \alpha_{-M m}$, while the
states in the upper group have $\alpha_{M m} \approx - \alpha_{-M
m}$. In each group states separate to pairs of symmetric and
antisymmetric states, as noted above. For large $\Ht$ the lower
level of each pair has $\alpha_{M m} \approx \alpha_{M -m}$ and the
higher level has $\alpha_{M m} \approx - \alpha_{M -m}$. Both the
energy spectrum and the form of the eigenstates discussed above
should be revealed in electromagnetic resonance experiments. In
Sec.~\ref{MR} we give predictions for such possible experiments, and
their relation to the calculated entanglement entropy.

\section{Low-T ENQI effective Hamiltonian}
 \label{Sec:transverse}

We now incorporate all the terms in (\ref{generalH}), with all
nuclear levels and the off-diagonal dipolar interactions, into the
full ENQI model for the $\LHx$ system, including all terms relevant
to the phase diagram at energies $> 10$mK.

We begin by dividing up the original Hamiltonian (\ref{generalH})
into the form
\begin{equation}
H = H_0 + H_1^{zz} + H_1^{\perp}
 \label{H-model}
\end{equation}
where
\begin{eqnarray}
H_0 &=& H_{\rm cf} + H_{\rm hyp}^{zz} + H_Z\nonumber \\
H_1^{zz} &=&  U_{dip}^{zz} \nonumber \\
H_1^{\perp} &=& H_{\rm hyp}^{\perp} + U_{\rm dip}^{\perp}
 \label{H01}
\end{eqnarray}
and where we have written the dipolar interaction in the form
\begin{equation}
U^{\alpha \beta}_{ij} = U^{zz}_{ij} + U^{\perp}_{ij}
 \label{Vz-Vx}
\end{equation}
with a non-diagonal term
\begin{eqnarray}
U^{\perp}_{ij} &=&
\frac{U_0}{J^2} {\cal R}_{ij}^{\perp} \nonumber \\
&=& \frac{U_0}{J^2}[{\cal R}_{ij}^{\alpha \beta}-{\cal R}_{ij}^{zz}]
 \label{V-perp}
\end{eqnarray}
where ${\cal R}_{ij}^{\alpha \beta}$ was defined in (\ref{UR}).

\begin{figure}
\psfrag{10^-12}{\hspace{0.0cm} \vspace{0.0cm} {\large $10^{-12}$}}
\psfrag{10^-10}{\hspace{0cm} \vspace{0cm} {\large $10^{-10}$}}
\psfrag{10^-8}{\hspace{-0.2cm} \vspace{0.0cm} {\large $10^{-8}$}}
\psfrag{10^-6}{\hspace{-0.2cm} \vspace{0cm} {\large $10^{-6}$}}
\psfrag{10^-4}{\hspace{-0.2cm} \vspace{0.0cm} {\large $10^{-4}$}}
\psfrag{10^-2}{\hspace{-0.2cm} \vspace{0cm} {\large $10^{-2}$}}
\psfrag{10^0}{\hspace{-0.38cm} \vspace{0.0cm} {\large $10^{0}$}}
\includegraphics[width =
\columnwidth]{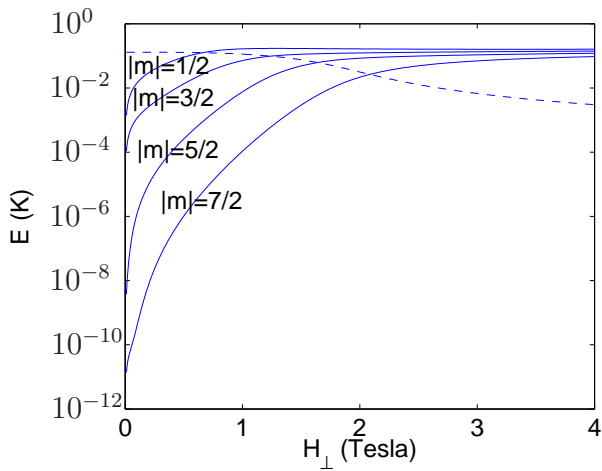}
 \caption{The quantum fluctuation amplitudes $\tilde{\Delta}_m$
induced by $\Ht$, plotted for $|m| =7/2,5/2,3/2,1/2$; the dashed
line shows the effective dipolar interaction $V_{ij}^{mm}$ at
x$=0.167$ for $|m| =7/2$ (the low-$T$ phase transition occurs when
this interaction $\sim \tilde{\Delta}_{7/2}$). Note how small is
$\tilde{\Delta}_m$ for large $\vert m \vert$ and small $\Ht$. }
    \label{figsplitting}
\end{figure}

In App.\ref{AppA} we derive a low energy effective Hamiltonian valid
for $T \ll \Omega_0$, $\mub \Ht \ll \Omega_0$. We do this in three
steps. We first derive the effective Hamiltonian for
$H=H_0+H_1^{zz}$, including Ising interactions terms only, and
obtain Eq.(\ref{Hdip2}). We then add $H_{\rm hyp}^{\perp}$ which
introduces a quantum term and obtain Eq.(\ref{mIsingHeffdip}). We
finally include $U_{\rm dip}^{\perp}$, which introduces an effective
random field\cite{SL06}, and an enhancement of the effective
transverse field\cite{SS05}, and as a final low-T effective
Hamiltonian for the $\LHx$ system we obtain
\begin{eqnarray}
H_{\rm eff} = &- \sum_{i,j,m,m'}
\tilde{V}_{im,jm'}^{zz}(\tilde{H}_i^{\perp},\tilde{H}_j^{\perp})\;
n_{im} n_{jm'} s_{im}^z
s_{jm'}^z \nonumber \\
&- \sum_{i,m} n_{im} \; [\epsilon_m +
\tilde{\Delta}_m(\tilde{H}_i^{\perp}) s_{im}^x] \nonumber \\
&+ \sum_i \gamma_i^z(\Ht)\sum_m n_{i,m} s_{im}^z .
 \label{mIsingHeff}
\end{eqnarray}
Here $\tilde{\Delta}_m$ are the effective transverse fields acting
on time reversed states with a given $|m|$, as defined in
Eq.~(\ref{pmstatesm}) (see Fig.\ref{figsplitting}), and $\gamma_i^z$
is an effective random field, defined in Eq.~(\ref{gammaj}). We note
explicitly the dependence of the interactions and the effective
transverse fields and random field on the site-dependent total
transverse field $\tilde{H}_i^{\perp}$ (\ref{tildeH}).

In the low-$T$ limit $kT \ll \omega_0$, and for $\Ht < \Htst(7/2)$,
we obtain
\begin{eqnarray}
H \;=\; - \sum_{i,j} \tilde{V}_{ij}^{zz}(\tilde{H}_i^{\perp},
\tilde{H}_j^{\perp}) \; s_i^z s_j^z \nonumber \\ \;-\; \sum_i
\tilde{\Delta}(\tilde{H}_i^{\perp})  s_i^x + \sum_i \gamma_i^z(\Ht)
s_i^z .
 \label{Heff-toyRF}
\end{eqnarray}
This Hamiltonian applies for any x$<1$, irrespective of what
thermodynamic phase results from it. Thus in $\LHx$ it is valid for
both the SG and FM regimes (the x dependence enters in the
interaction terms and in the effective fields). Note, that
$\tilde{\Delta}$ and $\gamma_i^z$ have a very different dependence
on $\Ht$ and dilution x. Thus, in the FM phase $\tilde{\Delta}$ and
$\gamma_i^z$ are independently tunable, by changing x and $\Ht$
\cite{Sch06}. In $\LHx$ one may thereby realize, for the first time,
both the quantum and the classical random field Ising models in a
ferromagnetic system (see the theoretical details in
Ref.\cite{Sch06}, and the experimental realization in
Ref.\cite{SBB+07}).

We emphasize the essential role played here by the nuclear spins.
They block quantum fluctuations. This is especially important for
the $\LHx$ system, whose peculiar crystal field Hamiltonian allows
electronic tunneling at second order in $\Ht$. If we drop the
nuclear spins we can get erroneous results (eg., that the effective
random field must come at the expense of appreciable quantum
fluctuations\cite{TGK+06}). For some purposes one can circumvent a
proper treatment of the hyperfine interactions by considering a
simplified crystal field Hamiltonian\cite{SL06,SSL07,Sch06} (see
also \cite{TVG08}), where tunneling between the electronic spins is
in high order perturbation. This gives the correct effective random
field, and the re-entrance of the cross-over $\Ht$ as a function of
dilution (see Sec.\ref{sec:R-NDdip}). However, for other purposes a
proper treatment of the hyperfine interactions is essential - eg.,
for the temperature and field dependence of the phase diagram (see
Sec.\ref{Sec:offD}), and for all of the dynamic properties.

\section{Mean field treatment of the Phase diagram}
 \label{Sec:offD}

The phase diagram of Quantum Ising systems like $\LHx$ has been the
object of extensive study for over three decades, and it was
realized early on that strong hyperfine interactions might be
important\cite{And73}. In the case of $\LHx$, for x$=1$ the phase
diagram was calculated in mean-field\cite{BRA96}, including the
longitudinal hyperfine interactions; this gave an enhancement of the
critical transverse field at low temperatures.

In this section we analyze the phase diagram of the $\LHx$ system in
various dilutions, where disorder effects have to be accounted for.
We first discuss the phase diagram of a model Hamiltonian, including
first the longitudinal dipolar and hyperfine interactions, and then
adding the transverse hyperfine interaction. Using this model we
make predictions for the behavior of the phase diagram and of the
magnetization of a general anisotropic dipolar system where the
conditions $A_0,V_0 \ll \Omega_0$ are well satisfied, for an
arbitrary ratio of $A_0/V_0$. This analysis also pinpoints the basic
physics dictating the phase diagram in the $\LHx$ system. However,
in the $\LHx$ system the condition $A_0 \ll \Omega_0$ is not that
well maintained. To solve for the phase diagram of the $\LHx$ system
we use exact diagonalization of the single Ho ion, a mean field
approximation for the inter-Ho interactions, and we take into
account the enhancement of the effective transverse field by the
transverse dipolar terms. We shall see that this then gives very
accurate results for the phase lines for $\LHx$ when x$ = 0.167$,
and we make predictions for x$ = 0.045$.

Finally, we discuss the nature of the phases at low $T$; this is
currently rather controversial. The hyperfine interactions again
play a central role, in reducing quantum fluctuations and slowing
the relaxation of the system to equilibrium in the low-$T$ quantum
regime. At finite transverse field we discuss the effect of the
effective longitudinal random field, emerging from the applied
transverse field.

\subsection{Classical Ising limit}
 \label{sec:ENQI-Toy-PD}

As shown above, if $\Ht$ is small, the transverse hyperfine
interactions play a minor role (the $\tilde{\Delta}_m$ are small),
and the only effect of the longitudinal hyperfine interactions is to
give a rather strong renormalization of the longitudinal dipolar
interaction between the Ising doublet spins $\tau_j^z$. The problem
in this Ising limit (neglecting the transverse terms) was studied
previously\cite{BD01}, but only for x$=1$. We give a treatment here
for all dilutions, and we also assume that $A_0$ is arbitrary -
surprisingly, the hyperfine interactions cannot be neglected even
when $A_0 \ll V_0$.

\subsubsection{Strong hyperfine interactions}
  \label{sec:toyPhase}

\begin{figure}
\psfrag{x}{\hspace{0.2cm} \vspace{0.5cm} {\large $T$}}
\psfrag{y}{\hspace{-1.3cm} \vspace{0cm} {\large $\Delta_0$}}
\includegraphics[width = 8cm]{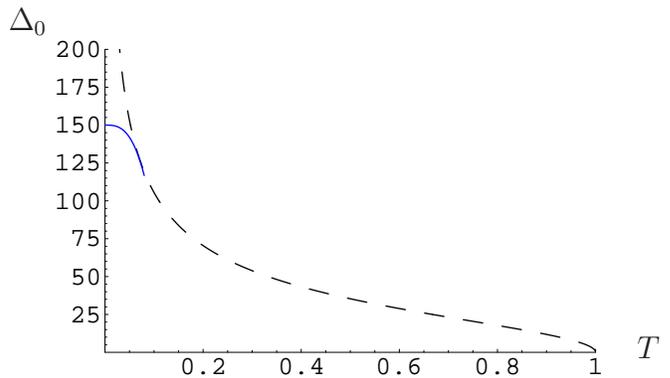}
\caption{Plot of the phase line separating the ordered and the
paramagnetic phases for the model (\ref{hfIsing}), in the regime
where $A_0 \gg V_0$, assuming $V_0=1, A_0=35$, and $\Omega_0=100$ in
arbitrary units. The dashed line shows Eq.(\ref{tctoy}), plotted
neglecting transverse hyperfine interactions; we see that $\Delta_c$
diverges as $T \rightarrow 0$. The blue solid line describes the
deviation from the classical Ising model at high transverse fields,
where the transverse hyperfine interaction are significant, giving a
QPT at $\Delta_0 \approx \Omega_0$ (see Sec.\ref{susec:transHyp}). }
    \label{figtoy}
\end{figure}

When $A_0 > V_0$, and $kT < \omega_0$, the relevant Hilbert space
comprises the lowest two electronuclear Ising-like levels, and we
consider the Hamiltonian (\ref{hfIsing}), which reduces to the
classical Ising Hamiltonian $H_{\rm eff}^{\parallel}$ given in
(\ref{ToyClassH}). The only effect of $H_{\perp}$ is to renormalize
$V_{ij}^{zz}$ to $\tilde{V}_{ij}^{zz} = \eta^2 V_{ij}^{zz}$. We can
then immediately deduce the whole phase diagram of the system. Since
$\eta$ is a function only of $\Delta_0/A_0$ [see Eq.~(\ref{eta2})]
we can write all expressions for the phase diagram in terms of
$\Delta_0$ instead of the actual transverse field $\Ht$. The
transition line as a function of $\Delta_0$ is shown in Fig.
\ref{figtoy} for $A_0 \gg V_0$ (ie., for x$ \ll 1$ in the $\LHx$
system); this diagram simply depicts the relation\cite{SS05}
\begin{equation}
T_c(\Delta_0) = \eta(\Delta_0)^2 T_c(0) \, .
\label{tctoy}
\end{equation}
If we now define $\epsilon \equiv (T_c - T)/T_c$ one finds that for
$\Delta_0/A_0, \epsilon \ll 1 $ (ie., small $H_{\perp}$ and $T \sim
T_c$) the phase transition line $\Delta_c(T)$ obeys the relation
\begin{equation}
\Delta_c = A_0 \sqrt{\epsilon} \, .
 \label{Eqepsilon}
\end{equation}
At $T=T_c/2$ one finds that $\Delta_c \approx A_0$. When
$\Delta_0/A_0 \gg 1$ there is still a finite remnant polarization of
the spin, and an ordered state at $T=0$; the transition line obeys
the relation $\Delta_c=A_0 \sqrt{(V_0/T)}$. This is quite different
from the TFQI model (\ref{Q-Ising}), where for $\Delta_0 > V_0$ the
system becomes a paramagnet, and a $T=0$ quantum critical point is
observed.

\vspace{3mm}

\subsubsection{renormalized Ising model for arbitrary $A_0/V_0$}
 \label{Sec:MF}

We now relax the condition $V_0 \ll A_0$,  so that all hyperfine
levels have to be included (however we still assume nuclear spin
flips are blocked).

For this case one can treat the Hamiltonian (\ref{hfIsing-m}) using
mean field theory; it then reduces to the mean field effective
Hamiltonian
\begin{equation}
H_{\rm MF} = \sum_i (hI_i^z - H_i^z) \tau_i^z - \sum_i \Delta_0
\tau_i^x
 \label{BD}
\end{equation}
where the site-dependent mean field is
\begin{equation}
H_i^z=\sum_j V_{ij}^{zz} \langle \tau_j^z \rangle \,
 \label{Hi}
\end{equation}
and $h \equiv \omega_0$.

Since (\ref{hfIsing-m}) is equivalent to the classical Ising
Hamiltonian (\ref{Hdip2}), the mean field Hamiltonian (\ref{BD}) is
equivalent to the mean field version of (\ref{Hdip2}), given by
\begin{equation}
H_{\rm MF}^{\parallel} \;=\; \sum_{im} n_{im} (\epsilon_{im} +
E_{im}\hat{s}_{im})
 \label{H_MF}
\end{equation}
where now the mean field is
\begin{equation}
E_{im} \;=\; \sum_{jm'} n_{jm'}\tilde{V}_{im,jm'}^{zz}
\langle\hat{s}_{jm'}^z \rangle .
 \label{Him}
\end{equation}

The mean field theory in the form (\ref{BD}) was solved some time
ago\cite{BD01} for the homogeneous case (where $\langle \tau_j^z
\rangle$ is independent of $j$, ie., the mean field is the same at
all sites), and applied to the ferromagnetic ${\rm LiHoF_4}$ system
(ie., when x$=1$).

In this section we extend this mean field approach to cover all
values of x, including the spin-glass regime, by allowing the local
mean field to vary from site to site. In order to allow easy
comparison with the previous work\cite{BD01}, we do this starting
from the Hamiltonian in the form (\ref{BD}) rather than
(\ref{H_MF}). An explicit derivation, given in App. \ref{AppB},
results in the self consistent equation

\begin{widetext}

\begin{equation}
1=\frac{\sum_m \frac{\Delta_0^2 V_0}{(h^2m^2+\Delta_0^2)^{3/2}}
\sinh{(\beta \sqrt{h^2m^2+\Delta_0^2})} + \frac{\beta
V_0h^2m^2}{h^2m^2+\Delta_0^2} \cosh{(\beta
\sqrt{h^2m^2+\Delta_0^2})}}{\sum_m \cosh{(\beta
\sqrt{h^2m^2+\Delta_0^2})}} \, . \label{BDlong}
\end{equation}

\end{widetext}
\noindent When $A_0 \gg V_0$ the solution of this equation
reproduces the results of Sec. \ref{sec:ENQI-Toy-PD}, and in
particular the phase diagram in Fig. \ref{figtoy}. Let us now
consider the regime $V_0 \gg A_0$. Expanding Eq.(\ref{BDlong}) in
small $\Delta_0$ one obtains the behavior of the transition line
$\Delta_c(T)$ near $T_c(0)$:
\begin{equation}
\Delta_c (\epsilon) = V_0 \sqrt{\epsilon} \, ,
\end{equation}
where $\epsilon \equiv (T_c - T)/T_c$. Because the dipolar
interaction now dominates, the results of the TFQI model are
reproduced near $T_c$; and at $T_c/2$ we find $\Delta_c \approx
V_0$. Surprisingly however, the hyperfine interaction, although
small, dictates the physics at low temperatures. Below a crossover
temperature $T^*=A_0^2/V_0$ one reaches a regime where $\Delta_c
> V_0$, and then $\Delta_c(T)$ is given by
\begin{equation}
\Delta_c(T) =A_0 \sqrt{(V_0/T)} \;\;\;\;\;\;(T<T^*)\,
 \label{largeVsmallT}
\end{equation}
corresponding to a transition temperature $T_c(H_{\perp}) = V_0
A_0^2/\Delta_0^2$. These are the exact same formulae found above for
the case $A_0 \gg V_0$. Thus, when $\Delta_0 > V_0$ the system gains
more energy from fluctuations than it does from the interaction.
However, since $\Ht$ cannot flip nuclear spins, a small remnant
magnetization proportional to $A_0/\Delta_0$ allows ordering at low
temperatures.

\begin{figure}
\begin{center}
\psfrag{x}{\hspace{0.2cm} \vspace{0.5cm} {\large $\Delta_0$}}
\psfrag{y}{\hspace{-1.3cm} \vspace{0cm} {\large $M_z$}}
\includegraphics[width = 8cm]{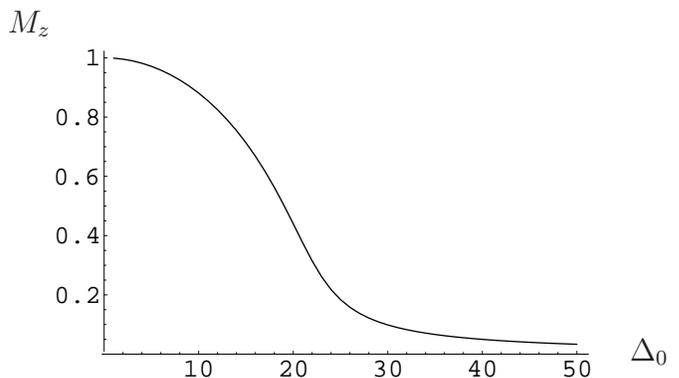}
\caption{The magnetization as function of transverse effective field
$\Delta_0$ is plotted for the Hamiltonian (\ref{BD}) with $h \equiv
2A_0/7 =1, V_0=20$. Note that for $\Delta_0 > V_0$ a remnant
magnetization of magnitude $h/\Delta_0$ is present. }
    \label{magnetizationBD}
\end{center}
\end{figure}

\begin{figure}
\begin{center}
\psfrag{x}{\hspace{0.2cm} \vspace{0.5cm} {\large $T$}}
\psfrag{y}{\hspace{-1.3cm} \vspace{0cm} {\large $\Delta_0$}}
\subfigure[]{\includegraphics[width = 6cm]{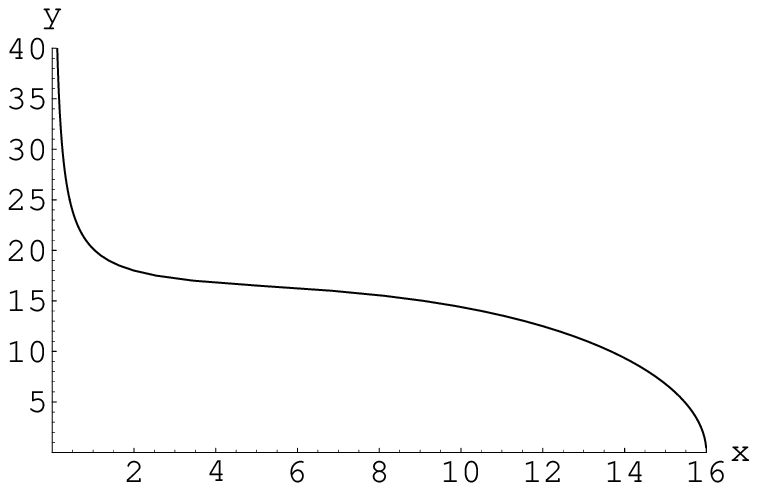}}
\subfigure[]{\includegraphics[width = 6cm]{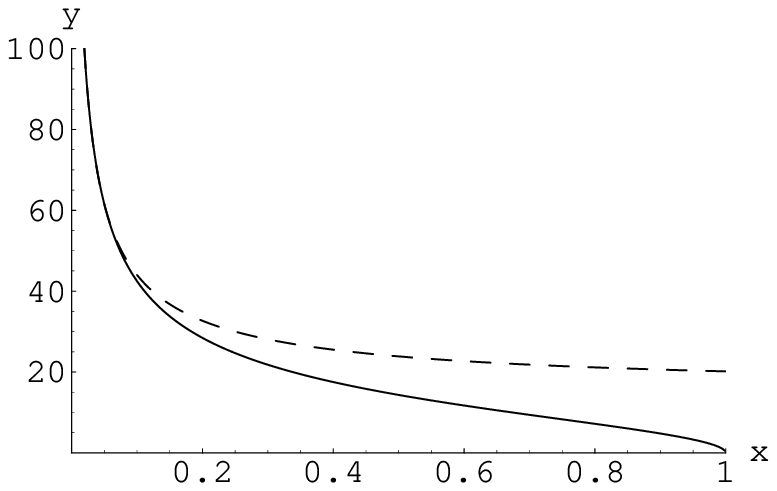}}
\caption{(a) The phase diagram of the Hamiltonian (\ref{BD}) for the
regime $A_0 \ll V_0$, with $h \equiv 2 A_0/7 = 1, V_0=16$.  (b)
Focusing on low temperatures, we compare the phase line in (a)
(dashed line) to a system with $A_0 \gg V_0$. We take $h=4, V_0=1$
to have the same value for $A_0 \sqrt{V_0}$, so that the low
temperature divergence of the critical field is the same for the two
systems. }
    \label{figphasedBD}
\end{center}
\end{figure}

In this mean field theory one thus finds 2 regimes, The first, when
$V_0 \gg A_0$ and $\Delta_0 \ll V_0$, is the standard Ising picture:
at $\Delta_0 =0$ the spins are in either state $\up$ or $\down$, and
the electronic degrees of freedom order. For finite $\Delta_0 \ll
V_0$ the spins fluctuate to the excited state at energy $V_0$.
However, when $\Delta_0 \gg V_0$ and/or in the whole parameter
regime for $A_0 \gg V_0$ the physical picture is different: the
relevant single Ho Ising states are the electro-nuclear states $\mid
\Uparrow \rangle, \mid \Downarrow \rangle$, Eq.(\ref{pmstates}), and
the phase transition line is dictated by their $H_{\perp}$ dependent
interaction, as discussed in Sec. \ref{sec:Hyp-perp}.

These two physical pictures are best illustrated by the value of the
magnetization at $T=0$. For $A_0 \gg V_0$, $M_z \propto \eta$, given
in Eq.(\ref{eta}). For $V_0 \gg A_0$ and $\Delta_0 \ll V_0$,
expanding Eq.(\ref{Eqmz}) in $\Delta_0/V_0$, one sees that $M_z = 1
- \Delta_0^2/(2 V_0^2)$, showing that the excitation energy is
$V_0$. However, when $\Delta_0 \geq V_0$, the hyperfine energy
dictates the magnetization, which is given by $M_z \approx A/V_0$
for $\Delta_0 = V_0$ and $M^z = A_0/\Delta_0$ for $\Delta_0 \gg V_0$
(see Fig. \ref{magnetizationBD}).

In Fig.\ref{figphasedBD} we plot the phase diagram of the mean field
Hamiltonian (\ref{BD}) as a function of $T$ and $\Ht$ for $V_0 \gg
A_0$. In the low $T$ regime one can compare this with the phase
diagram of a system with $A_0 \gg V_0$ and a similar value of $A_0
\sqrt{V_0}$. As expected from (\ref{largeVsmallT}), for $T \ll T^*$
the two systems have the same behavior.

In Ref. \cite{BD01} a similar phase diagram was calculated for $\LH$
and compared to experiment\cite{BRA96}. This comparison was made by
rescaling the theoretical curve to agree with the experiments at the
lowest temperature. However the condition $H_{\perp} \ll
\Omega_0/\mub$ is then not well satisfied at criticality, and the
transverse hyperfine interactions are important. By forcing the
theory and experiment to coincide in the regime where the theory is
not applicable, a discrepancy with experiment over the whole
temperature range is obtained (see Fig. 1(b) in Ref.\cite{BD01}).
This can be corrected for $T > 0.1$K by choosing the scaling
parameter better. However, in order to obtain a good fit with the
experimental phase diagram at the lowest temperatures one has to
take into account the transverse hyperfine terms\cite{BRA96}. For
x$<1$ the off-diagonal dipolar interactions have to be included as
well. These interactions are considered next.

\subsection{Effect of transverse hyperfine interaction}
 \label{susec:transHyp}

Independent of the ratio $A_0/V_0$, for $\Delta_0 \gg A_0,V_0$ we
found for $H_{\rm hyp}^{\perp}=0$, which is equivalent to $\Omega_0
\rightarrow \infty$, that $\Delta_c=A_0 \sqrt{(V_0/T)}$, diverging
as $T \rightarrow 0$. This pathology arises because we need to
include the transverse hyperfine terms. With $H_{\rm hyp}^{\perp}
\neq 0$ the splitting $\tilde{\Delta}$ becomes appreciable for $\mub
H_{\perp} \approx \Delta_0 \approx \Omega_0$ (in this regime $\mub
H_{\perp} \approx \Delta_0$), while $V_{eff}=V_0A_0^2/\Delta_0^2 \ll
A_0$. Thus, for $A_0,V_0 \ll \Omega_0$ a quantum phase transition is
obtained within the regime of the applicability of the Hamiltonian
(\ref{Heff-toy}). The divergence of $\Delta_c$ is rounded, as we
schematically draw in solid line in Fig. \ref{figtoy}. Similar
rounding off of $\Delta_c$ occurs for $V_0 \gg A_0$, Fig.
\ref{figphasedBD}.

As mentioned above, the condition $A_0 \ll \Omega_0$ is not that
well satisfied in the $\LHx$ system. Still, at low x, where $A_0 \gg
V_0$, the phase transition occurs within the regime of applicability
of the Hamiltonian (\ref{Heff-toy}); recall that in
Fig.\ref{figsplitting} we plot $V_{eff} \propto \langle J_z \rangle
^2$ for x$=0.167$, taking $V_{eff}(H_{\perp}=0)$ to equal the value
of $T_c=0.13$K. The value of $\Ht$ where $\tilde{\Delta} \approx
V_{eff}$ is smaller than $\Htst$. This is true for all smaller
dilutions x as well. At x$=0.167$ one expects the quantum phase
transition to occur at $\Ht \approx 2$T, where $\tilde{\Delta}
\approx V_{eff}$. Thus, three energy scales govern the phase
transition. The spin-spin interaction $V_0$ dictates $T_c$ at zero
field, the hyperfine interaction $A_0$ dictates the phase diagram at
finite $\Ht$, and the larger anisotropy scale $\Omega_0$ dictates
the position of the quantum critical point, since quantum
fluctuations only become important when $\upt \mub H_{\perp} \ex
\approx \Omega_0$. It is for this reason\cite{SS05} that in $\LHx$
it is much easier to disorder the ordered phase thermally, rather
than quantum mechanically\cite{WBRA93}, specially when x$ \ll 1$.

\begin{figure}[ht!]
\subfigure[]{\includegraphics[width =
\columnwidth]{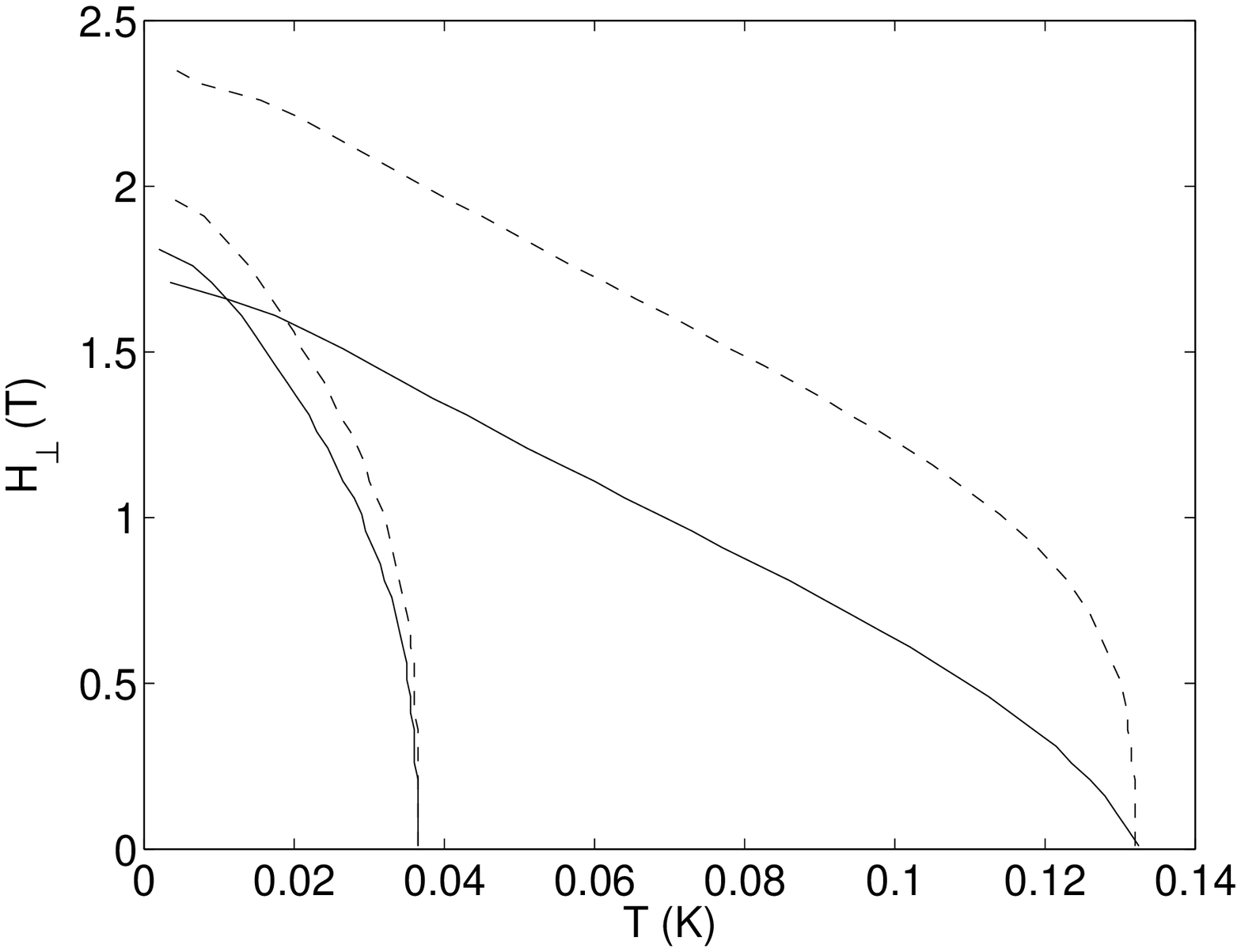}}
\subfigure[]{\includegraphics[width =
\columnwidth]{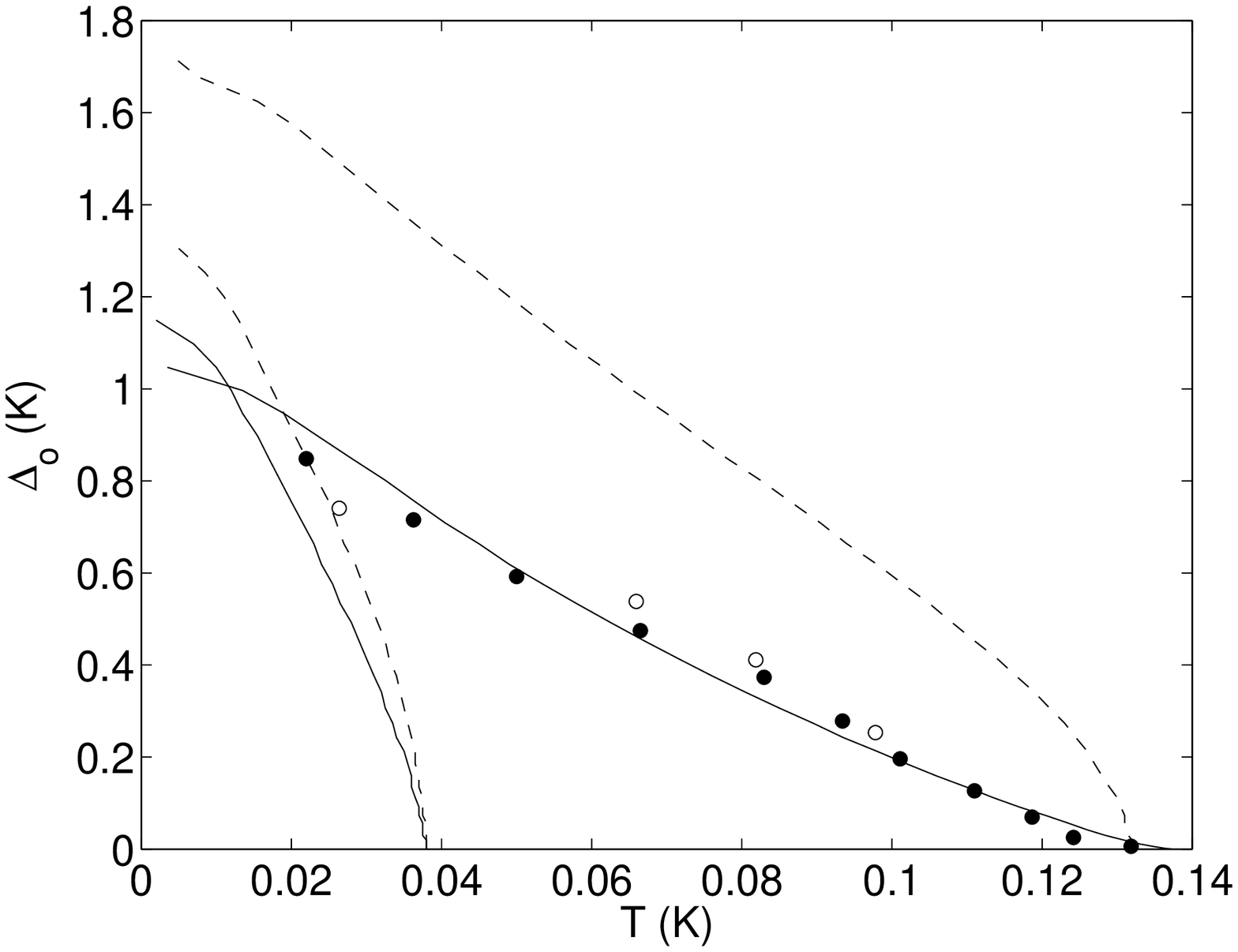}}
 \caption{(a) The phase diagram for x$=0.167$ (thick lines) and x$=0.045$
(thin lines) as a function of $H_{\perp}$ and $T$. The solid
(dashed) lines are calculated with (without) the offdiagonal dipolar
interactions. The low $T, H_{\perp}=0$ phase is believed to be a
spin glass for both dilutions, with a crossover at finite $\Ht$
between a quasi spin-glass and a paramagnet. Note the "re-entrant"
behavior predicted when the off-diagonal dipolar interactions are
included (as seen by the crossing of the phase lines at $\Ht \sim
1.65~T$); see text. (b) The same phase diagram as in (a) now plotted
as a function of $\Delta_0,T$, to allow comparison with
experiment\cite{WBRA93} at x$=0.167$. Filled and empty circles,
taken from Fig.1 of Ref.~\cite{WBRA93}, denote the PM-SG crossover
from dynamical measurements and nonlinear susceptibility
respectively. With the inclusion of the off-diagonal dipolar
interactions, good quantitative agreement is obtained.}
    \label{figphased}
\end{figure}

For $x \ll 1$ one can calculate the phase diagram for $\LHx$
including all hyperfine terms using the effective Hamiltonian
derived in Sec.~\ref{Sec:transverse}. However as we have seen, at
higher temperatures this Hamiltonian breaks down at quite low
transverse fields, because of the mixing of higher levels (cf. Fig.
\ref{figsplitting}). Therefore, for larger x, where the dipolar
interactions are stronger, quantum criticality occurs at $\Ht >
\Htst$, where all the nuclear levels are well mixed. This is the
case for x$=1$ where a quantum phase transition is observed at
$4.9$T\cite{BRA96}.

We therefore adopt a different approach, which covers all values of
x, and calculate the phase diagram numerically, including both
$H_{\rm hyp}^{\|}$ and $H_{\rm hyp}^{\perp}$, starting from the
Hamiltonian
\begin{eqnarray}
H_{\rm F} \;\;=\;\; \sum_j  (H_j^{\rm cf} &-& g_J \mub
H_{\perp}J_j^x \nonumber \\
&+& A_J \vec{I_j} \cdot \vec{J_j} + \sum_i U_{ij}^{zz} J_i^z J_j^z)
\, ,
 \label{longdipH}
\end{eqnarray}
in which the single spin Hamiltonian is exact; and we then treat the
interactions in mean field approximation, i.e., we assume
\begin{equation}
\sum_{i} U_{ij}^{zz} J_i^z J_j^z \rightarrow  U_{\rm MF} \langle
J_j^z \rangle J_j^z ,
 \label{MFT2}
\end{equation}
(see also App.~\ref{AppB}). One of the central results of this paper
is that the single atom Hamiltonian dictates much of the physics of
$\LHx$, and we shall see that the mean field approximation to the
interactions has only a small effect on the results.

The phase diagram for x$=0.167$ and x$=0.045$ is drawn in dashed
lines in Fig~\ref{figphased}. Comparing this calculation for
x$=0.167$ with experiment, we see it naturally explains why it is
much harder to disorder the spin-glass phase quantum mechanically
than thermally. Going to x=$0.045$, we see the reduction in $T_c$ is
$\propto $ x, while the reduction in $H_{\perp}^c$ is much smaller,
as can be anticipated from the requirement $V_{eff} \approx
\Delta_0$ (Fig. \ref{figsplitting}).

However, the agreement with experiment is still not perfect for
small x. For x$=0.167$ one obtains a larger critical field at $T=0$
and a qualitatively different behavior near $T_c(0)$. As was
discussed in Ref.\cite{SS05}, these differences can not be
attributed to the mean field approximation, but testify to the
inadequacy of the Hamiltonian (\ref{longdipH}). This is since the
behavior near $T_c(0)$ should follow Eq.(\ref{Eqepsilon})(for
x$=0.167$ the condition $V_0 \ll A_0$ is well satisfied), and the
values for $\Delta_c$ obtained at the lower temperatures in the
experiment\cite{WBRA93} necessitate the existence of appreciable
quantum fluctuations at $H_{\perp} \approx 1T$, which contradicts
the results shown in Fig.\ref{figsplitting}. To explain things we
now finally turn to the non-diagonal dipolar terms.

\subsection{Random non-diagonal dipolar terms}
 \label{sec:R-NDdip}

To account for the experimental phase diagram, one has to include
the dependence of the effective field on the offdiagonal terms of
the dipolar interaction\cite{SS05}. These add an effective random
longitudinal field, and in the spin-glass regime also enhance the
effective transverse magnetic field, as is explained in
App.~\ref{AppA}, Sec.~\ref{ENQI-dip}. The random longitudinal field
is crucial in dictating the nature of the phase at finite $\Ht$, as
it destroys long-range SG order\cite{SL06,SSL07}; however, at least
for x$\ll 1$ it does not strongly affect the position of the phase
line, because (i) the effective random longitudinal field is zero at
$\Ht=0$, and is small for $\Ht \ll \Omega_0/\mub$; (ii) it is random
in sign, with only a small effect on the typical interaction; and
(iii) at large $\Ht$, where $\gamma_i^z$ is appreciable, the
crossover to the paramagnetic phase depends only weakly on $V_0$, as
can be inferred from Fig.\ref{figsplitting}.

Thus, in calculating the phase diagram we neglect the random
longitudinal fields, and consider only the enhancement of the
effective transverse field by the off-diagonal dipolar interactions.
This enhancement depends on x and $\Ht$; here we follow
Ref.\cite{SS05} in neglecting the dependence on $\Ht$. We further
assume that this enhancement is proportional to x when x$\ll 1$,
i.e. we write a total transverse mean field $\tilde{H}_{\perp} =
H_{\perp} + H_{\perp}^d$, with $H_{\perp}^d \propto $ x. Note that
this mean field is just the average of the transverse field
$\tilde{H}_i^{\perp}$ that we discussed in section
Sec.~\ref{ENQI-dip}., ie., $H_{\perp}^d = \langle
\tilde{H}_i^{\perp} \rangle$.

This leads to a satisfying quantitative agreement with the
experimental phase diagram at x$=0.167$ (see Fig.\ref{figphased}).
$T_c(0) \propto V_0 \propto $ x, while $H_{\perp}^c$ at the $T=0$
transition depends mainly on the energy scale $\Omega_0$. For x$\leq
0.167$, $H_{\perp}^c$ should therefore change only slightly with
dilution. The dilution dependence of $H_{\perp}^c$ is a result of
two effects. First, since the transition occurs when $\tilde{\Delta}
\approx V_0$, there is a slow decrease of $H_{\perp}^c$ with x (slow
because $\tilde{\Delta}$ varies rapidly with $\Ht$; see
Fig.\ref{figsplitting}). Second, $H_{\perp}^d \propto $ x, and
giving a further reduction $\propto $ x in $H_{\perp}^c$. This leads
to the interesting prediction that for low enough x there will be an
increase of $H_{\perp}^c$ with decreasing x, so that $H_{\perp}^c$
has a minimum at some x; this is seen in our figure by the crossing
of the phase lines (see Fig.\ref{figphased}). In analogy with the
re-entrant behavior one sees in some systems on variation of an
external field, we can call this a prediction of a kind of
're-entrance' as a function of concentration x.

It is interesting that the combined effect of the hyperfine
interactions and the transverse dipolar interactions leads to this
re-entrant behavior. Even though the effect of the transverse
dipolar interactions is only a weak effect compared to that of the
hyperfine terms, it is just enough to tip the system into
re-entrance. Note however that without the much stronger hyperfine
effect on the phase diagram, this would not have happened. We remark
again that we do not think that it is possible to explain the phase
diagram without incorporating the hyperfine terms (eg., by including
only dipolar interactions\cite{TGK+06}; cf our discussion in section
\ref{Sec:transverse}).

\subsection{Nature of the low temperature phase}
 \label{nature}

As we have seen it is possible to derive an accurate phase diagram
without saying too much about the nature of the phases themselves.
In fact the nature of the low-$T$ phases of $\LHx$ has been rather
controversial in recent years. Here we would like to outline several
rather important implications of our results. We divide our
discussion between the zero transverse field case and the case of
finite $\Ht$.

\subsubsection{Zero Transverse field}
 \label{sec:ZTF}

At all dilutions, the $\LHx$ system is paramagnetic at high
temperatures. However, as mentioned above, at low temperatures the
phase of the system is dilution dependent. It is well established
both experimentally\cite{REY+90} and theoretically\cite{BH07} that
for x$> {\rm x}_F$ the system orders ferromagnetically at low
temperatures, where values for ${\rm x}_F$ are in the range
$0.2-0.5$. However, at low dilutions the nature of the phase is
controversial. Theoretically, it is argued that a spin-glass phase
should exist at all dilutions x$ \ll 1$\cite{SA81}. Experimentally,
it was argued that at x$=0.167$ the system has a low temperature
glass phase\cite{WBRA93} while for x$=0.045$ the
experiment\cite{GPRA02} revealed a very intriguing and yet
unexplained behavior of the imaginary part of the susceptibility, in
which its width in the frequency domain {\it narrows} as temperature
is lowered, and therefore received the name ``anti spin-glass''.
Recently, however, these results were challenged by Jonsson et al.,
\cite{JMW+07}, who claim for x$=0.167,0.045$ that there is no phase
transition to the spin-glass phase. Furthermore, their analysis
suggests that the system at the above two dilutions exhibits similar
characteristics. A similar controversy arose regarding the specific
heat of the system and its consequences regarding the nature of the
phase at x$=0.045$\cite{GRAC03,QMG+07}. Note, that it is difficult
to reach equilibrium conditions both experimentally, near the
transition\cite{JMW+07,Bar07}, and numerically, using Monte
Carlo\cite{BH07}, and therefore further studies will be useful in
resolving the low temperature phase of the diluted $\LHx$.

Our analysis above does not depend on the precise nature of the
ordered phase, and therefore can not lead to definite conclusions
regarding this question. However, since the Hamiltonian
(\ref{generalH}) gives a comprehensive description of the system
down to a few mK, some clarifying statements based on our analysis
can be made.

(i) the only differences between the $\LHx$ compounds at x$=0.167$
and x$=0.045$ are the strength of the dipolar interaction, both in
the magnitude of the typical terms and in the distribution due to
randomness. All the single Ho properties, which, as discussed above,
dictate much of the physics, stay unchanged. Thus, we have every
reason to believe that at x$=0.045$ the equilibrium low temperature
phase is also a spin-glass. However, as is shown in
Fig.\ref{figphased}, its $T_c(0)$ is reduced to roughly $35mK$, and
according to this the experiments at this
dilution\cite{GPRA02,GRAC03,QMG+07,JMW+07} were done in the
paramagnetic regime.

(ii) as we show above, the dynamics of the system at low
temperatures is significantly slowed down by the coupling to the
nuclear spins (see also Ref.\cite{JMW+07}). Indeed, the peculiar
features in the spin susceptibility at x$=0.045$\cite{GPRA02} were
obtained as temperature was reduced to below $150$mK. At this
temperature the higher nuclear spin levels start to be depleted, and
all but few of the Ho atoms are in either state $\a$ or $\abar$.
Thus, the system can not take advantage of the much faster
transitions between the higher nuclear spin states (see
Fig.\ref{figsplitting}), and the dynamics slow down appreciably. The
data of Quilliam et al.\cite{QMG+07}, showing that the peak in the
specific heat occurs in a similar temperature for
x$=0.02,0.045,0.08$ supports the view that single spin physics, and
in particular the hyperfine interactions, are significant in the
interpretation of the experiments in these dilutions.

(iii) In Ref. \cite{GRAC03} it was argued that for x$=0.045$ the
internal transverse field resulting from the offdiagonal terms of
the dipolar interaction stabilize a low temperature spin liquid
state. It was further argued there that this is correct also for
transverse fields which are effectively reduced by a factor of
$10^4$. The analysis of Ref. \cite{GRAC03} was done in the
electronic degrees of freedom. However, in the regime relevant to
the experiment\cite{GPRA02} the effective Hamiltonian
(\ref{Heff-toy}) is valid, with zero random longitudinal field.
Therefore, the analysis should be done considering the
electronuclear degrees of freedom, within the framework of the
Hamiltonian (\ref{Heff-toy}). In particular, the effective
transverse field due to the offdiagonal dipolar interactions at
$\Ht=0$ is much smaller than the values considered in
Ref.\cite{GRAC03}, as can be inferred from the log scale graph in
Fig.\ref{figsplitting}.

\subsubsection{Finite Transverse Field}
 \label{sec:FTF}

Turning now to non-zero $\Ht$, we note first that there is a crucial
difference between the FM phase and the SG phase, if the latter
exists at $\Ht = 0$. In the FM regime, the lower critical dimension
$d_c = 2$ (cf. ref. \cite{IM75}), and in $3$D the FM phase is stable
to a small random field. Thus we expect that the FM phase, which
exists for large x, will survive at finite $\Ht$.

However, if one supposes that for intermediate x one has a SG phase
at $\Ht = 0$, then the critical dimension is $d_c = \infty$ (cf.
refs.\cite{FH86,FH88}), and so the long-range SG order should be
destroyed by an infinitesimal random field\cite{SL06,SSL07}. As is
well-known, this means that the system will no longer be a
homogeneous SG, but instead domains of finite size will be created -
each one will have internal SG order but the order will be
uncorrelated between different domains. The correlation length
$\xi$, which is essentially the domain size, is given
by\cite{SL06,SSL07}
\begin{equation}
\xi \approx \left(\frac{\Omega_0}{\mub \Ht}\right)^\frac{1}{(3/2) -
\theta_d} \, .
 \label{corl}
\end{equation}
where $\theta_d \approx 0.2$ is the stiffness
exponent\cite{FH86,FH88}. Essentially the system is able to gain
energy from the random field by creating domains. Referring to
Eq.(\ref{E2}), we see that this energy gain is a result of the two
terms in the numerator contributing with the same sign, i.e. an
effective enhancement of the transverse magnetic field.

\section{Experimental consequences}
 \label{experiments}

In this paper we have derived new results regarding the single
particle properties of the Ho ion in the $\LHx$ systems, as well as
the phase diagram. We have also addressed the regime where $A_0 \ll
V_0$, which is not applicable to the $\LHx$ system but is the more
abundant regime in general. In this section we address the relation
between our results and possible experiments.

\subsection{Single spin properties}

A central result of this paper is the derivation of the low energy
effective Hamiltonian for the $\LHx$ system, as a generalized Ising
Hamiltonian in the electronuclear degrees of freedom
(\ref{mIsingHeff}). This effective Hamiltonian is completely
determined by the effective random fields $\gamma_i^z$, and the
single ion parameters $\epsilon_m$, $\tilde{\Delta}_m$, and $\eta_m$
[see Eq.~(\ref{eta-m})], the latter determining the effective spin
and therefore the effective spin-spin interaction. Below we suggest
magnetic resonance and $\mu$SR experiments that can measure the
single ion parameters directly, and verify the mechanism leading to
the enhancement of the effective transverse field and the emergence
of an effective random longitudinal field. With regard to magnetic
resonance experiments, we give explicit quantitative predictions for
the Rabi frequency of excitations to various levels. We interpret
these predictions in terms of the calculated entanglement entropy of
the ground state as function of transverse field, in agreement with
our analysis in Sec.~\ref{sec:singleHo}.

\subsubsection{Magnetic Resonance Experiments}
\label{MR}

\begin{figure}[ht!]
\subfigure[]{\includegraphics[width =
\columnwidth]{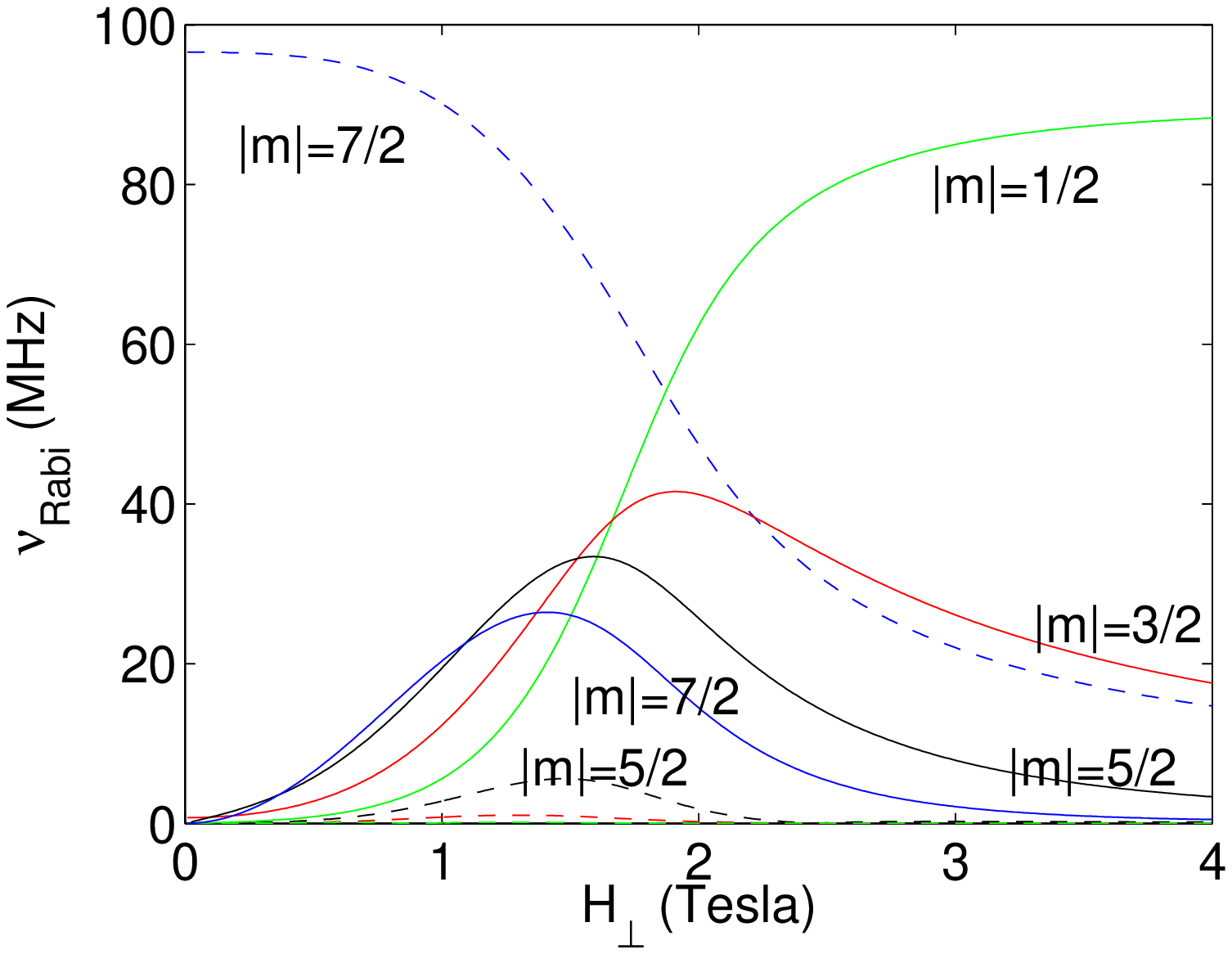}}
\subfigure[]{\includegraphics[width =
\columnwidth]{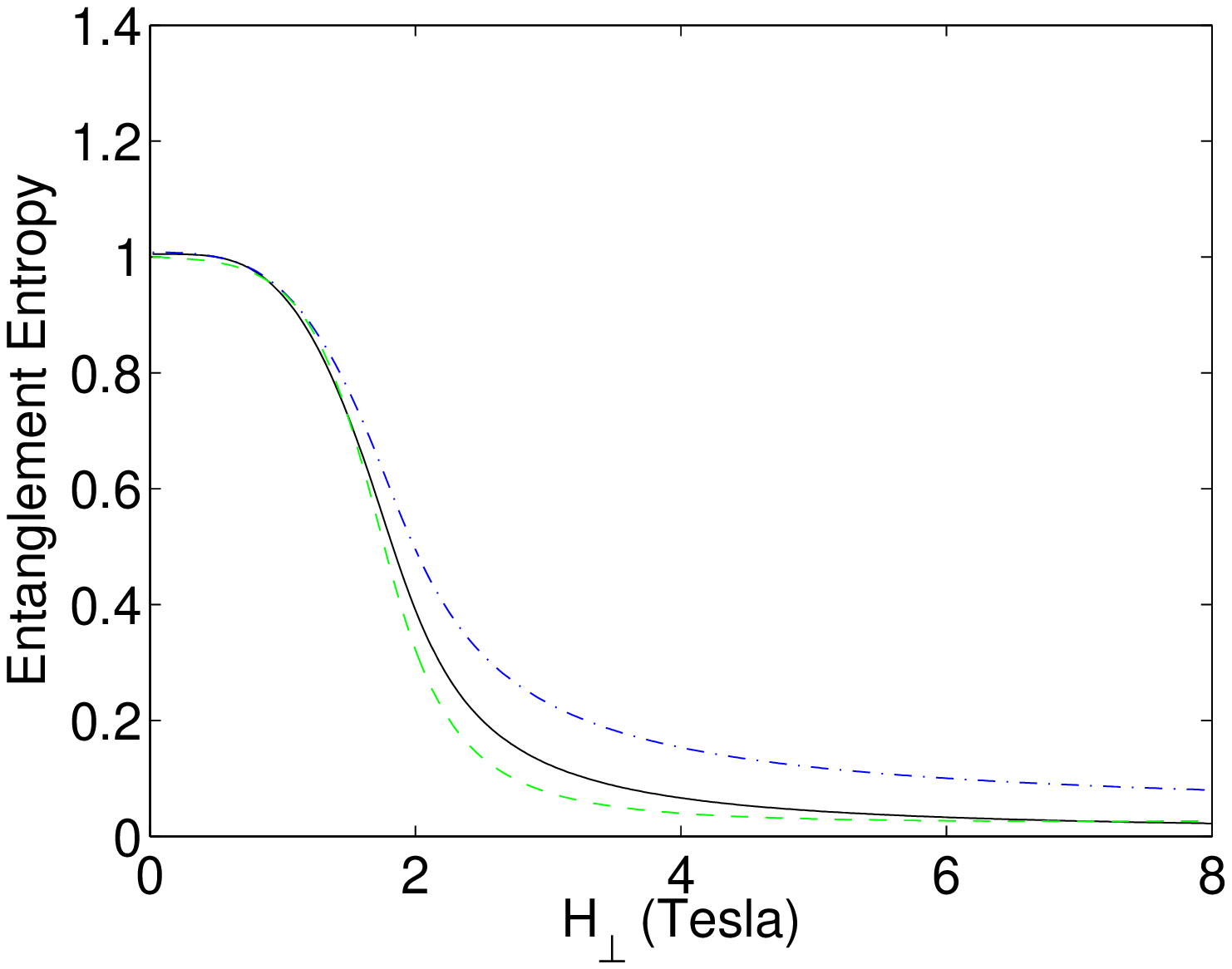}}
 \caption{(a) Rabi frequencies of magnetic resonance transitions to
 electro-nuclear levels within the space of the electronic Ising
 doublet, for oscillating field along $\hat{z}$ with amplitude $1m$T, as
function of a transverse magnetic field. All transitions are to
anti-symmetric states (dashed blue levels in Fig.\ref{fig16levels}).
Dashed (solid) lines correspond to levels within the lower (upper)
$8$ states. (b) Entanglement entropy of the ground state as function
of $\Ht$ (solid black line). Plotted for comparison are the
normalized Rabi frequency to level $2$ (first excited state, blue
dot dash line) and the subtraction from unity of the normalized Rabi
frequency to level $9$ (green dashed line). }
    \label{NMRESR}
\end{figure}

One obvious way of probing the low energy properties of the $\LHx$
system is via magnetic resonance experiments. Specifically, such
experiments can be used to quantify $\epsilon_m$,
$\tilde{\Delta}_m$, and the nature of the wave functions as function
of $\Ht$. In Fig.~\ref{NMRESR}(a) we plot the Rabi frequencies,
given by
\begin{equation}
\nu_{Rabi} = \frac{2 b_z \mid \langle 1 \mid g_J \mub J_z + g_N
\mu_N I_z \mid 2 \rangle \mid}{h} ,
\end{equation}
for a magnetic resonance transitions between the ground state and
excited states as function of transverse field $H_x$, for an AC
field along $\hat{z}$. Only the lowest $16$ levels are considered.
From symmetry, only transitions to antisymmetric states (plotted
dashed in Fig.\ref{fig16levels}, see also discussion in
Sec.~\ref{sec:singleHo}) are possible. For $\Ht \rightarrow 0$ the
only allowed transition is to the first excited state, in agreement
with the form of the electronuclear states in Eq.~(\ref{pmstates}),
in terms of which the effective Hamiltonian (\ref{mIsingH-eff}) is
written. Finite Rabi frequency to other states is allowed at $H_x
\neq 0$, consequence of the mixing of the states (\ref{pmstates})
resulting from $H_{\rm hyp}^{\perp}$, and is thus larger for larger
$|m|$. As discussed in Sec.\ref{sec:singleHo}, with increasing $H_x$
the electronic and nuclear spins disentangle, and the electronic
state at high fields is approximately the symmetric state $(\up +
\down)/\sqrt{2}$ for the lower $8$ states and the antisymmetric
state $(\up - \down)/\sqrt{2}$ for the upper $8$ states. For this
reason, except for the first excited state at small $H_x$, the Rabi
frequency is larger to the states in the upper group. At large $H_x$
The Rabi frequency to state $9$ (the lowest level in the upper
group) dominates, in agreement with the picture (see
Sec.~\ref{sec:singleHo}) that levels in the lower and upper groups
of $8$ have similar nuclear states, respectively.

The disentanglement of the electronic and nuclear states can be
quantified by calculating, as function of $H_x$, the entanglement
entropy $-Tr(\rho_{_I} \log{\rho_{_I}})$, where

\begin{equation}
\rho_{_I} \equiv \sum_M \langle M \mid g.s. \rangle \langle g.s.
\mid M \rangle \label{EE}
\end{equation}
is the reduced density matrix in the subsystem of the nuclear spin.
The entanglement entropy is shown for the ground state in solid line
in Fig.~\ref{NMRESR}(b). In dashed line we re-plot the Rabi
frequency to the first excited state scaled to $1$ at $H_x=0$. In
dot-dash line we plot $1 - \tilde{\nu}_{\rm Rabi}(9)$, where
$\tilde{\nu}_{\rm Rabi}(9)$ is the scaled Rabi frequency to level
$9$. Although not exact, we see that the diminishing of $\nu_{\rm
Rabi}(2)$ and the emergence of $\nu_{\rm Rabi}(9)$ with increasing
field is a measure of the (dis)entanglement of the electronic and
nuclear spins. A naive conclusion from the above would be that for
large $\Ht$ the nuclear spins decouple from the electronic spins,
and therefore the effective Hamiltonian (\ref{Q-Ising}) is
recovered. The fact that for x$=1$ the soft mode is gapped near the
quantum phase transition (at $4.9$T)\cite{RPJ+05,RJP+07} suggests
that this simplified model is not suitable also in this regime.

\begin{figure}[ht!]
{\includegraphics[width = \columnwidth]{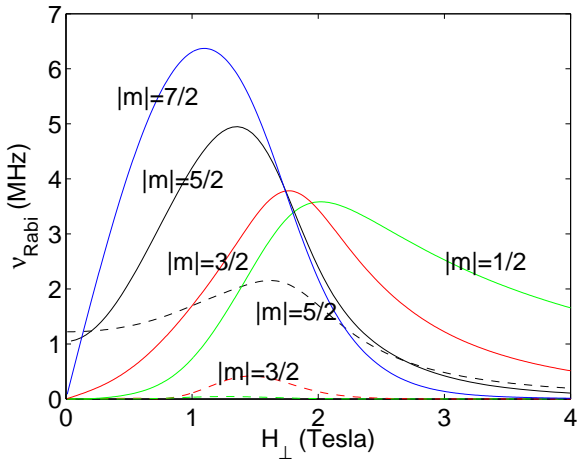}}
 \caption{Rabi frequency of magnetic resonance transitions for an AC
 field in the $x$ direction. Details are as in Fig.\ref{NMRESR}, only
 here the transitions are to symmetric states
 (solid green levels in Fig.\ref{fig16levels}).}
    \label{NMRESRx}
\end{figure}

In Fig.\ref{NMRESRx} we plot the Rabi frequency as function of
$H_x$, for an AC field in the x direction. The relevant matrix
element is then $\langle 1 \mid g_J \mub J_x + g_N \mu_N I_x \mid 2
\rangle$. The operator $J_x$ changes the $z$ component of the
electronic spin. The relevant matrix element is then proportional to
the amplitude of $\ex$ in the ground state. This amplitude,
resulting from the transverse hyperfine interaction, is small, $\sim
O(A_0/\Omega_0) \approx 10^{-2}$, and at $H_x=0$ is finite only for
the state with $\mid I_z \mid = 5/2$. This is why the intensity for
a longitudinal AC field is so much larger than that for a transverse
AC field (compare Fig.~\ref{NMRESR}(a) and Fig.~\ref{NMRESRx}). For
an AC field along $\hat{z}$, only transitions to symmetric states
(plotted solid in Fig.\ref{fig16levels}) are possible.

From the picture of the system without the transverse hyperfine
interaction, where the $16$ states at zero field are eigenstates of
$I_z$, one might expect that for a transverse AC field the dominant
intensity would come from the nuclear operator. Surprisingly, it is
the electronic operator that dominates the magnetic resonance
experiment. This is because $A_0/\Omega_0 \gg \mu_{\rm N}/\mub$.
Thus, although the levels are predominantly nuclear spin levels, the
relevant experiment is basically an ESR experiment.

All the results above are valid for single Ho ions, and can be
checked in very dilute samples, where interactions are negligible.
For larger x, in the SG regime, the interplay of the offdiagonal
dipolar interactions and the applied transverse field results in an
effective enhancement of the transverse field and the emergence of
an effective random field\cite{SS05,SL06} (see details in
App.~\ref{AppA}, Sec.~\ref{ENQI-dip}). This result, shown in
Secs.~\ref{sec:R-NDdip},~\ref{sec:FTF} to be crucial for the
structure of the phase diagram in the SG regime\cite{SS05} as well
as for the nature of the phase itself\cite{SL06,SSL07}, can actually
be verified by measuring e.g. $\nu_{\rm Rabi}(2)$ as function of
$\Ht$ for different x. The offdiagonal dipolar interactions should
lead not only to a dispersion in $\tilde{\Delta}_{7/2}$, but also to
an x dependent shift upward of its mean value. This shift can be
checked against our approximation in Sec.~\ref{sec:R-NDdip}.

\vspace{3mm}

\subsubsection{$\mu$SR Experiments}

According to Eq.~(\ref{eta}), the single Ho spin moment at low
temperatures $\propto \eta$, and so decreases with $\Ht$. Such a
field dependence of the individual magnetic moments could be
directly measured using $\mu$SR. The magnetic field at the muon site
is proportional to the magnetic moment size of the material, which,
in a diluted sample, is given by the nearest Ho ion. Such a
measurement should be done at dilution x$\ll 1$, both because our
prediction is for the regime where $A_0 > V_0$, and because then the
contribution from more distant Ho ions will be smaller.

\subsection{$\LHx$ Phase diagram}

In Sec.\ref{sec:R-NDdip} We have given explicit quantitative
predictions for the phase diagram of the $\LHx$ system as function
of x. Our results for x=$0.0167$ are in good agreement with
experiment, and our predictions regarding the phase diagram at
x$=0.045$, and the re-entrant cross-over field as function of
dilution, can be checked experimentally in a straight forward way.
However, here we would like to suggest an experiment that would
directly probe the significance of the hyperfine interactions in
dictating the phase diagram of the $\LHx$ system at low x.

We use the above result, predicting that at $H_x \approx 2-3$T, the
magnetic resonance intensities for transitions to levels with $\mid
I_z \mid = 1/2, 3/2$ are appreciable. Thus, one could in principle,
by populating these states, change the critical field at low-T: a
non-equilibrium occupation of these excited electro-nuclear levels
would lead to stronger quantum fluctuations (cf. Fig.
\ref{figsplitting}), and therefore to a lower critical field. This
opens up the rather fascinating possibility of controlling the
quasi-equilibrium phase diagram of the system by driving a
steady-state non-equilibrium nuclear spin population.

\subsection{Limitations of the $\LHx$ system}

The $\LHx$ compound is a particularly useful test system: It is a
well defined Ising system, with a doubly degenerate ground state,
quantum fluctuation are easily tunable at moderate transverse
fields, and x can be varied over a huge range. However, there are at
least two limitations on this system, viz.:

(i) The allowed values of the dipolar spin-spin interaction $V_0$,
hyperfine interaction $A_0$, and crystal anisotropy energy
$\Omega_0$, do not test the whole parameter range. Thus, eg., to
observe our prediction that the hyperfine coupling dictates a
diverging $H_c$ at low $T$ for either $A_0 \ll V_0$ or $A_0 \gg V_0$
given that $A_0,V_0 \ll \Omega_0$ (see
Figs.\ref{figtoy},\ref{figphasedBD}), we need a system where the
latter condition is well satisfied.

(ii) For $\Ht=0$, the ground state is degenerate. Inducing quantum
fluctuations coupling the two ground states requires a transverse
field. However, the application of $\Ht$ results in an emerging
random field. As a result, the quantum phase transition between the
SG and FM phases can not be seen as function of $\Ht$, but only as
function of a parameter that does not break time reversal symmetry,
e.g. pressure\cite{SL06}. To observe such a transition one would
need a system where quantum fluctuations between the Ising ground
states are appreciable at $\Ht = 0$. One would then have to tune the
dilution so that at ambient pressure the typical spin spin
interactions is of the order of the quantum fluctuations, and look
for the transition as function of pressure.

\section{Conclusions}
 \label{sec:conc}

In this paper we have considered anisotropic quantum magnetic
systems in which both dipolar and hyperfine interactions play a
role. We have shown that the transverse field Ising model is not
sufficient to describe such systems; instead, we have given a
theoretical treatment of an Electronuclear Quantum Ising model which
can do the job. The hyperfine interactions set the scale for the
field at the quantum critical point, even in systems in which the
hyperfine interaction is weaker than the dipolar spin-spin
interaction. We have given a detailed treatment of the $\LHx$
compound, calculating the phase diagram for all dilutions x, and
giving explicit numerical results for x$ = 0.045$ and x$ = 0.167$.
We explain the experimental result that thermal fluctuations more
easily destabilize the ordered phase than quantum mechanical
fluctuations. Off-diagonal dipolar interaction terms are shown to
reduce the transverse critical field $H_c^{\perp}$, and a prediction
for a non-monotonic critical field as a function of x is given.
Experimental consequences of our results as well as possible
measurements of the parameters of the effective Hamiltonian are
discussed.

We note that our results have wider implications in 2 ways, which
will be explored elsewhere. First, as just noted, they can be
applied to many other dipolar quantum magnets. Second, the nuclear
spins will clearly have an even more profound effect on the
dynamical properties of these systems than on the phase diagram -
indeed, the big surprise is quite how important they are for the
thermodynamics, even when $A_0 \ll V_0$.

{\it Note added in proof} -- Our prediction for the reentrance of
the crossover field as function of dilution was coincidentally and
independently discovered experimentally by Ancona-Torres et.
al.\cite{ASAR08}.

\vspace{3mm}

It is a pleasure to thank G. Aeppli, B. Barbara, B. Malkin. A.
Morello, and J. Rodriguez, for useful discussions. This work was
supported by NSERC in Canada, and by PITP.

\appendix

\section{Derivation of the effective Hamiltonian} \label{AppA}

In this Appendix we derive the effective Hamiltonian in
Eq.~({\ref{mIsingHeff}).

\subsection{Ising-like terms}
 \label{sec:Hyp-perp}

We are interested in the $2I+1$ lowest levels, shown for the $16$
levels of the Ho ion in $\LHx$ in Fig.\ref{fignuclearsplitting}. The
term $H_1^{\perp}$ in Eq.(\ref{H01}), has zero matrix element
between any of these $16$ low energy states, and thus must mix
higher crystal field states. This results in contributions of order
$V_0/\Omega_0, A_J/\Omega_0 \ll 1$. We therefore begin by discussing
in this subsection the longitudinal interaction terms, i.e. the
Hamiltonian $H_l=H_0 + H_1^{zz}$.

In the subspace of the lowest $2I+1$ electronuclear states, we have
seen that $H_l$ reduces to

\begin{equation}
H_{\rm el} = \sum_i \omega_0 \tau_i^z I_i^z  - \sum_{i,j}
V^{zz}_{ij} \tau_i^z \tau_j^z  - \sum_i \Delta_0 \tau_i^x .
 \label{hfIsing-m}
\end{equation}

Let us first discuss this Hamiltonian for $T \ll \omega_0$. Then,
one can simplify the model to include only the levels with $I_z =
\pm I$ (ie. with $I_z = \pm 7/2$ for $\LHx$), described in the
previous section. The effective Hamiltonian then reduces
to\cite{SS05}:
\begin{equation}
H_{\rm le} = 2 A_0 \sum_i \tau_i^z \sigma_i^z  - \sum_{i,j}
V^{zz}_{ij} \tau_i^z \tau_j^z  - \sum_i \Delta_0 \tau_i^x
 \label{hfIsing}
\end{equation}
where $\tau_i$ acts upon the two electronic states $\up,\down$ at
site $i$, $\sigma_i$ acts upon the two nuclear spin states with $I_z
= \pm 7/2$, and the coupling $A_0 = I\omega_0 \sim 0.7~K$.

For $H_{\perp} \neq 0$ the two low energy Ising doublet states are
given by\cite{SS05}
\begin{eqnarray}
\mid \Uparrow \rangle &=& c_1 \mid a \rangle + c_2 \mid b \rangle
\nonumber
\\
\mid \Downarrow \rangle &=& c_1 \mid \bar{a} \rangle + c_2 \mid
\bar{b} \rangle \; \; ,
 \label{pmstates}
\end{eqnarray}
where $\mid a \rangle, \mid b \rangle, \mid \bar{a} \rangle, \mid
\bar{b} \rangle$ are defined in Sec.~\ref{sec:hyperfine}, and
\begin{eqnarray}
c_1 &=& \alpha \Delta_0 \;; \;\;\;\;\; c_2 \;=\; \alpha
[A_0-\sqrt{A_0^2+\Delta_0^2}] \\
\alpha &=& [\Delta_0^2 +(A_0-\sqrt{A_0^2+\Delta_0^2})^2]^{-1/2} .
 \label{c12}
\end{eqnarray}
Thus, as noted just above, the longitudinal part of the hyperfine
interaction blocks quantum fluctuations between the relevant Ising
states.

It then follows that a transverse field $\Ht \ll \Omega_0$ can only
renormalize the effective spin of what is just a {\it classical}
Ising system: one finds
\begin{equation}
\langle \tau_{\pm}^z(H_{\perp}) \rangle \;=\; \eta \langle
\tau_{\pm}^z(0) \rangle \hspace{1cm} ; \hspace{1cm} \eta = (c_1^2 -
c_2^2) \, ,
 \label{eta}
\end{equation}
with $\langle \tau_{-}^z \rangle = - \langle \tau_{+}^z \rangle$.
Note that
\begin{eqnarray}
\eta=1-\frac{\Delta_0^2}{2A_0^2} &\;& (\Delta_0 \ll A_0) \nonumber
\\
\eta=A_0/\Delta_0 &\;& (\Delta_0/A_0 \gg 1)
 \label{eta2}
\end{eqnarray}
Absorbing this renormalization into the dipolar interaction, $H_{\rm
le}$ (\ref{hfIsing}) reduces to:
\begin{eqnarray}
H_{\rm eff}^{\parallel} &=& - \sum_{i,j} \tilde{V}_{ij}^{zz} s_i^z
s_j^z
\;\;\;\;\;\; (H_{\perp} \ll \Omega_0/\mub) \\
\tilde{V}_{ij}^{zz} &=& \eta^2 V_{ij}^{zz}
 \label{ToyClassH}
\end{eqnarray}
where $\hat{s}_j$ is a spin-half matrix operating on the states
$\mid \Uparrow \rangle$ and $\mid \Downarrow \rangle$ of the $j$-th
spin, such that $\hat{s}_j^z \mid \Uparrow \rangle = \mid \Uparrow
\rangle$, and $\hat{s}_j^z \mid \Downarrow \rangle = -\mid
\Downarrow \rangle$, etc.

Thus we have shown the equivalence of the two Hamiltonians $H_{\rm
le}$ in (\ref{hfIsing}) and $H_{\rm eff}^{\parallel}$ in
(\ref{ToyClassH}), and both are applicable in the low-$T$ limit $kT
\ll \omega_0$, when $\Ht \ll \Omega_0$.

For higher temperatures, $\omega_0 \lesssim T \ll \Omega_0$, the
Hamiltonian (\ref{hfIsing-m}) has to be considered. Following the
arguments above, only levels with the same $I_z$ mix (see
Fig.~\ref{fignuclearsplitting}), and the generalization of
Eqs.(\ref{pmstates})-(\ref{c12}) results in

\begin{eqnarray}
\mid \Uparrow, m \rangle &=& c_{1m} \mid \uparrow, m \rangle +
c_{2m}
\mid \downarrow, m \rangle   \nonumber\\
\mid \Downarrow, -m \rangle &=& c_{1m} \mid \downarrow, -m \rangle +
c_{2m} \mid \uparrow, -m \rangle \; \; ,
 \label{pmstatesm}
\end{eqnarray}
with coefficients
\begin{eqnarray}
c_{1m} &=& \alpha_m \Delta_0 ; \nonumber \\
c_{2m} &=& \alpha_m [m\omega_0+\sqrt{m^2\omega_0^2+ \Delta_0^2} ]
 \label{c12m}
\end{eqnarray}
and
\begin{equation}
 \alpha_m = \left[\Delta_0^2 +
(m\omega_0+\sqrt{m^2\omega_0^2+\Delta_0^2})^2\right]^{-1/2} .
 \label{am}
\end{equation}
If we choose $m = -I$, these equations revert to
(\ref{pmstates})-(\ref{c12}) for the 2 lowest levels.

We now define a set of pseudospin-$1/2$ degrees of freedom $\{
\hat{s}_{im} \}$, operating in the subspace spanned by the pair of
degenerate levels $\vert \Uparrow, m \rangle$, $\vert \Downarrow, -m
\rangle$. In terms of these pseudospins we then obtain, for the
Hamiltonian $H_l$ defined above [Eq.~(\ref{hfIsing-m})], the
renormalized effective Hamiltonian
\begin{eqnarray}
H_{\rm eff}^{\parallel} &=& \sum_{jm} n_{jm} \epsilon_m(\Ht) \nonumber \\
&-& \sum_{i,j;m,m'}n_{im} n_{jm'} \tilde{V}_{im,jm'}^{zz}(\Ht)
s_{im}^z s_{jm'}^z
 \label{Hdip2}
\end{eqnarray}
where we have defined a psedospin occupation number $n_{jm}$ such
that $\sum_m n_{jm} = 1$, and defined energies
\begin{equation}
\epsilon_m(\Ht) = sgn(m) \sqrt{(m^2 \omega_0^2 + \Delta_0^2)} ,
 \label{epsilonm}
\end{equation}
\begin{equation}
\tilde{V}_{im,jm'}^{zz} = \eta_m \eta_{m'} V_{ij}^{zz}
 \label{V-mm'}
\end{equation}
and a renormalisation factor
\begin{equation}
\eta_m = (c_{1m}^2 - c_{2m}^2) .
 \label{eta-m}
\end{equation}
Thus again we obtain a generalized classical Ising model, now
involving the entire set $\{ \hat{s}_{im} \}$ of pseudospins. There
are $2I+1$ pseudospins per site (only one of which is occupied at
any time), interacting with pseudospins at the other sites.

We emphasize that the Hamiltonians (\ref{hfIsing-m}) and
(\ref{Hdip2}) are entirely equivalent. The great advantage of
(\ref{Hdip2}) (and its low-$T$ simplification in (\ref{ToyClassH}))
is that the physics is correctly displayed, that of a classical
Ising system; and this is done using the physically meaningful
energy scales for this regime.

\subsection{Transverse Hyperfine terms}
 \label{sec:Heff-Hyp}

Now suppose we neglect all dipolar interactions between the
electronic spins, but we now switch on the full hyperfine coupling,
including the transverse hyperfine term. The general effect of this
is seen in a plot (Fig. \ref{fig16levels}) of the 16 relevant
eigenenergies for the Ho ion in the $\LHx$ system, as a function of
$\Ht$.

As $\Ht$ increases, levels separate into 2 groups of eight, given by
symmetric and antisymmetric combinations of states with the same
$m$. Each pair of levels $\mid \Uparrow, m \rangle, \mid \Downarrow,
-m \rangle$ which is related by time reversal symmetry is then split
by the combination of $\Ht$ and $H_{\rm hyp}^{\perp}$.

If we wish to write an effective Hamiltonian for this system in the
original basis of $2I+1$ levels, we get a rather interesting result.
After truncating the full $H_{\rm cf}$ down to the 2 lowest
electronic levels, we can write for a {\it single ion} (ignoring now
all interactions between ions):
\begin{equation}
H_i^{\rm eff} \;=\; -{1 \over 2} \sum_j \Delta_0 [\tau_j^+
e^{i\hat{\bf A}[{\bf I}_j]} + H.c.] +  \omega_0 \sum_i \tau_i^z
\hat{I}_i^z
 \label{H-SB}
\end{equation}
where the matrix element $U^{fi}_j = \langle f \vert e^{i\hat{\bf
A}[{\bf I}_j]} \vert i \rangle$ is defined between an initial state
$\vert i \rangle = \vert \uparrow j; \chi_i({\bf I}_j) \rangle$
before the electronic spin flips, and a final state $\vert f \rangle
= \vert \downarrow j; \chi_f({\bf I}_j) \rangle$ after it flips,
involving some initial and final nuclear spin wave-functions
$\chi_i({\bf I}_j)$ and $\chi_f({\bf I}_j)$. The operator
$\hat{U}^{(j)}$ is defined as
\begin{equation}
\hat{U}_j =  e^{i\hat{\bf A}[{\bf I}_j]} \;\equiv \;   \exp {{i
\over \hbar} \int^{\Downarrow}_{\Uparrow} dt H_{\rm
hyp}^{\perp}({\bf I}_j,t)}
 \label{U-inter}
\end{equation}
and in general the $(2I+1)\times(2I+1)$-matrix $\hat{\bf A}[{\bf
I}]$, which operates in the Hilbert space of ${\bf I}_j$, causes
transitions between different hyperfine levels when the central spin
(here the Ho spin) flips. A calculation of $\hat{\bf A}[{\bf I}]$ is
actually quite lengthy, since typically it involves multiple
transitions between the nuclear spin states. Thus the Hamiltonian
(\ref{H-SB}) is in general a rather complicated object, taking the
form of a $(2I+1)\times(2I+1)$-matrix acting on the lowest states of
the system. For $\LHx$ this means a $16 \times 16$ matrix, which is
quite unwieldy.

In the present paper we are only interested in the thermodynamic
properties of our system. This allows a considerable simplification,
which we can see most simply by rewriting the Hamiltonian
(\ref{H-SB}) for an isolated ion (ie., again ignoring dipolar
interactions) in terms of our pseudospins, in the following form:
\begin{equation}
H_i^{eff} = \sum_{m} H_m^o(\hat{s}_{im}) + \sum_{m \neq m'} \delta
\tilde{H}_{mm'}(\hat{s}_{im}, \hat{s}_{im'})
 \label{Hmm'}
\end{equation}
where the 'diagonal' terms have the matrix form
\begin{equation}
H_m^o = \left(
\begin{tabular}{cc}
$  \epsilon_m $ & $ \tilde{\Delta}_m $\\
$ \tilde{\Delta}_m^{\dagger} $ & $ \epsilon_m $
\end{tabular}
\right)
  \label{H_m}
\end{equation}
in the basis where $\hat{s}_{im}^z$ is diagonal, and the
non-diagonal terms couple different pseudospins.

We look first at the diagonal terms. The diagonal energies
$\epsilon_m$ {\bf (\ref{epsilonm})} are just the eigenvalues of
$H_0$ in (\ref{H01}); in zero field one has from (\ref{H-long}) that
$\epsilon_m \sim m\omega_0$ for $\LHx$, with splitting $\sim 205~mK$
between adjacent pairs of levels in the $Ho$ ion. At low $\Ht$ the
transition terms $\tilde{\Delta}_m$ split each pair of degenerate
$m$-states into the symmetric and antisymmetric combinations $\vert
\pm, m \rangle = (\vert \Uparrow, m \rangle \pm \vert \Downarrow, -m
\rangle/\sqrt{2}$. The $\tilde{\Delta}_m$ are just the quantum
fluctuation amplitudes between the eigenstates $\vert \Uparrow, m
\rangle$ and $\vert \Downarrow, -m \rangle$ of the classical Ising
system, induced by the transverse hyperfine coupling.

Now consider the non-diagonal term $\delta
\tilde{H}_{mm'}(\hat{s}_{im}, \hat{s}_{im'})$ in (\ref{Hmm'}). The
crucial point here is that when $\Ht/\Omega_0 \ll 1$, this term will
be unimportant for the phase diagram because a non-diagonal
couplings between different pseudospins $\hat{s}_{im},
\hat{s}_{im'}$ involves a sequence of $\vert m - m' \vert$ nuclear
flips. If we call these non-diagonal matrix elements
$\tilde{\Delta}_{mm'}$, then for $\Ht/\Omega_0 \ll 1$
$\tilde{\Delta}_{mm'} \ll \omega_0$ and so it can hardly affect the
level spacing or any other thermodynamic properties. Thus, as far as
the thermodynamics is concerned, we can get away with using the form
$H_m^o$ in (\ref{H_m}), when $\Ht/\Omega_0 \ll 1$.

Summarizing, including the transverse hyperfine terms the effective
low energy Hamiltonian for a single ion can be written as
\begin{equation}
H_i^{\rm eff} \;\sim\; \sum_{i,m} n_{im} [\epsilon_m(\Ht) -
\tilde{\Delta}_m(H_{\perp}) s_{im}^x]
 \label{Hmo-eff}
\end{equation}
where both $\epsilon_m$ and $\tilde{\Delta}_m$ depend on the
transverse field. If we now include back the longitudinal dipolar
interaction, we obtain a Hamiltonian
\begin{eqnarray}
H_{\rm eff} = &- \sum_{i,j,m,m'}
\tilde{V}_{im,jm'}^{zz}(H_{\perp},\omega_0) n_{im} n_{jm'} s_{im}^z
s_{jm'}^z \nonumber \\
&- \sum_{i,m} n_{im} [\epsilon_m + \tilde{\Delta}_m(H_{\perp},
\omega_0) s_{im}^x]
 \label{mIsingHeffdip}
\end{eqnarray}
which is the generalization of Eq.(\ref{Hdip2}) that now includes
quantum fluctuations.

At very low temperatures, $T \ll \omega_0$ (which, for x$\ll 1$,
include the whole phase diagram) only the $2$ lowest electro-nuclear
states are relevant, and the above Hamiltonian reduces to\cite{SS05}
\begin{equation}
H = - \sum_{i,j} \tilde{V}_{ij}^{zz}(H_{\perp}) s_i^z  s_j^z -
\tilde{\Delta}(H_{\perp}) \sum_i  s_i^x \, ,
 \label{Heff-toy}
\end{equation}
which is just Eq.(\ref{ToyClassH}) with the addition of quantum
fluctuations. and is the ENQI Hamiltonian given in
Eq.~(\ref{IsingH-eff}).

We have still not quite finished; we must finally add in transverse
dipolar terms, which introduce one further modification to the
effective Hamiltonian.

\subsection{Non-diagonal Dipolar Interactions}
 \label{ENQI-dip}

The 'non-diagonal' dipolar terms $U_{ij}^{\perp}$ couple the Ho
spins in higher order in the small parameter
$U^{\perp}_{ij}/\Omega_0$, and so they have typically been neglected
when discussing anisotropic dipolar systems, including $\LH$.
However, in Refs. \cite{SS05,SL06,SSL07} it was shown that terms
$\sim J_j^z J_i^x$ can be rather important. For x$=1$ these terms
cancel by symmetry, but not for $0 < x < 1$, where even when $\Ht =
0$ they induce quantum fluctuations\cite{GRAC03}. For $\Ht \neq 0$
these terms can enhance or reduce the quantum fluctuations induced
by the applied field. To quantify this effect let us consider the
regime $\Ht \ll \Omega_0$ and write the original Hamiltonian in
Eq.(\ref{generalH}) or Eq.(\ref{H-model}) as
\begin{equation}
H = H_{\rm long} + H_{\rm trans}
\end{equation}
where
\begin{equation}
H_{\rm long} = H_{\rm cf} + H_{\rm hyp}^{zz} + U_{\rm dip}^{zz}
\end{equation}
and
\begin{equation}
H_{\rm trans} = - \sum_i g_J \mub \Ht J^x + \sum_{ij} U_{ij}^{zx}
J^z J^x .
\end{equation}
Here we neglect $H_{\rm hyp}^{\perp}$, and all terms other than
$U_{ij}^{zx}$ in $U_{\rm dip}^{\perp}$ which do not contribute in
lowest order of perturbation theory\cite{SL06,SSL07}.

$H_{\rm long}$ has a low-T ordered phase, which, depending on the
dilution, is FM or a SG\cite{REY+90,Ros96,BH07}. Let us denote
either of the two symmetry broken ground states of the ordered phase
by $\psi_0$. Introducing $H_{\rm trans}$ as a perturbation lowers
the energy of $\psi_0$ by\cite{SL06,SSL07,Sch06}
\begin{equation}
 E_{\psi}^{(2)} =  - \frac{\bpsio ( \sum_{i \neq j} U_{ij}^{zx}
J_i^z  J_j^x + g_J \mub \Ht \sum_i J_i^x)^2 \psio}{\Omega_0} \, .
 \label{E2}
\end{equation}
Thus, the offdiagonal dipolar terms add to the applied transverse
field a term

\begin{equation}
H_x^{(r)}({\bf r}_i) = \frac{\sum_j U_{ij}^{zx} \langle J_i^z
\rangle}{g_J \mub} .
\end{equation}
This additional field is random in sign, and can enhance or decrease
the quantum fluctuations generated by the applied field $\Ht$. Since
enhancing quantum fluctuations reduces the energy of the system,
configurations in which $H_x^{(r)}$ is in the direction of the
applied field are energetically favorable. The terms in
Eq.~(\ref{E2}) proportional to $H_{\perp}^2$ and to $U^2$ are
independent of the spin configuration of the system. However, the
cross term depends on the location and orientation of the spins, and
results in an effective random longitudinal
field\cite{SL06,SSL07,Sch06} $\gamma_i^z(\Ht)$ given by

\begin{equation}
\gamma_i^z = \frac{2 \mub \Ht J^2}{\Omega_0} \sum_j U_{ji}^{zx} \;\;
= \;\; c_i \frac{2 \mub \Ht J^2}{\Omega_0}U_0 , \label{gammaj}
\end{equation}
where $J=\langle J_z \rangle \approx 5$ is the single spin magnetic
moment, $c_i$ is a random number with $\langle c_i \rangle = 0$ and
${\it Var}(c_i) \equiv c^2(\rm x)$ is dilution dependent. For $(1 -
{\rm x}) \ll 1$ we have $c({\rm x}) = c' (1 - {\rm x})$\cite{Sch06},
with $c' \approx 1$. Note that the effective random field exists
both in the SG\cite{SL06,SSL07} and in the FM\cite{Sch06} regimes.
The randomness is a result of the quenched disorder, so the
interplay between the applied transverse field and the off-diagonal
dipolar terms converts spatial disorder to randomness in the
effective longitudinal field.  Unlike the quantum fluctuation
amplitudes $\tilde{\Delta}_m$, this random field is independent of
$m$ to zeroth order in $\omega_0/\Omega_0$.

It is clear from the definitions of the additional transverse field
$H_x^{(r)}({\bf r}_i)$ that its actual values and distribution
depends on the particular configuration adopted by the $\{ s_{im}^z
\}$ in the phase of interest. Thus this final term in the effective
Hamiltonian actually depends on what state the system is in. In
general this extra field simply renormalizes the total field acting
on the system; we have a new transverse field
\begin{equation}
\tilde{H}_i^{\perp} = \Ht + H_x^{(r)}({\bf r}_i) .
 \label{tildeH}
\end{equation}
This means that all the parameters in the effective Hamiltonian that
formally depend on the transverse field must now depend on
$\tilde{H}_i^{\perp}$ rather than $\Ht$. This inevitably introduces
some randomness in these parameters, from one site to another.
However, as is discussed in Sec. \ref{nature}, in the SG regime, the
system forms finite size domains that maximize the energy gain from
the random field $\gamma_i^z$. This is actually done by having a
finite average value for $H_x^{(r)}({\bf r}_i)$ in the direction of
$\Ht$, thus increasing quantum fluctuations, and results in an
effective enhancement of the applied magnetic field. With the
addition of the effective random field and the effective enhancement
of the transverse field we obtain the effective Hamiltonian in
Eq.~({\ref{mIsingHeff}).

\section{Classical Ising limit in mean field}
\label{AppB}

In this Appendix we derive Eq.(\ref{BDlong}). Consider a given site
$i$, with local longitudinal field $H_i$. We assume that as we cross
the transition line into the spin-glass phase the expectation value
of each spin grows at the same rate, i.e. $\langle \tau_j^z \rangle
= \alpha_{ij} \langle \tau_i^z \rangle$. It then follows that
\begin{equation}
H_i = V_i \langle \tau_i^z \rangle
\end{equation}
with $V_i=\sum_j V_{ij}^{zz} \alpha_{ij}$. Now the partition
function becomes $Z= \prod_i Z_i$ with
\begin{equation}
Z_i = Tr \{exp{[-\beta[(hI_i^z-H_i^z)\tau_i^z - \Delta_0
\tau_i^x]]}\}
 \label{partition}
\end{equation}
and the average magnetization of spin $i$ is given by
\begin{equation}
\langle \tau_i^z \rangle = \frac{1}{Z_i} Tr \{\tau_i^z
exp{[-\beta[(hI_i^z-H_i^z)\tau_i^z - \Delta_0 \tau_i^x]]}\} \, .
 \label{mz}
\end{equation}
Equations (\ref{partition}) and (\ref{mz}) are generalizations of
Eqs.(6) and (10) of Ref. \onlinecite{BD01} to the case where $V_i$
is site dependent. Note that $V_i$ is positive by definition. Assume
now that the paramagnet-spin-glass phase transition occurs when the
mean field equation gives a finite magnetization for spins at sites
with some typical $V_i=V_0$. Then, defining $M_z \equiv \langle
\tau^z \rangle$ one obtains
\begin{equation}
M_z = \frac{\sum_m \frac{hm+V_0 M_z}{\bar{H}(m)} sinh[\beta
\bar{H}(m)]}{\sum_m cosh[\beta \bar{H}(m)]}
 \label{Eqmz}
\end{equation}
where $\bar{H}(m)=\sqrt{(hm+V_0 M_z)^2+\Delta_0^2}$ is the total
magnitude of the mean field. This equation allows us to obtain the
phase diagram for any ratio of $A_0/V_0$, keeping $H_{\perp} \ll
\Omega_0/\mub$. Near the transition line (so $M^z \ll 1$) and
following Banerjee and Dattagupta\cite{BD01}, we expand
Eq.(\ref{Eqmz}) in $M^z$ and obtain Eq.(\ref{BDlong}).

\end{document}